\documentclass[fleqn,usenatbib,useAMS]{mnras}
\usepackage{graphicx}
\usepackage{amsmath} 
\usepackage{amssymb}
\usepackage[]{url}
\usepackage{multirow}
\usepackage{widetext}
\usepackage{lineno} 
\usepackage{acronym}
\usepackage{multirow}
\usepackage{tabularx}
\usepackage{ulem} 
\renewcommand{\emph}[1]{\textit{#1}}

\usepackage{booktabs} 

\def \prd {PRD}
\def \apj {ApJ}
\def \apjs {ApJS}
\def \apjl {ApJL}
\def \mnras {MNRAS}
\def \aap {A\&A}

\def \nat {Nature}

\def\deg{$^\circ$}

\def\t9{T$_{90}$}

\def\deg{$^{\circ}$}

\newcommand{\Rate}{\mathcal{R}}

\newcommand{\rate}{$\mathrm{Gpc}^{-3}\, \mathrm{yr}^{-1} $\,}
\newcommand{\eflux}{$ \mathrm{erg}\, \mathrm{s}^{-1}\, \mathrm{cm}^{-2} $\,}
\newcommand{\pflux}{$ \mathrm{ph}\, \mathrm{s}^{-1}\, \mathrm{cm}^{-2} $\,}
\newcommand{\lum}{$\,\mathrm{erg}\, \mathrm{s}^{-1} $\,}

\newacro{DQ shift}{Data Quality shift}
\newacro{FAP}{false alarm probability}
\newacro{EPO}{Education and Public Outreach}
\newacro{OURCA}{Office of Undergraduate Research and Creative Activity }
\newacro{DQ}{Data Quality}
\newacro{LIGO}{Laser Interferometer Gravitational Wave Observatory}
\newacro{aLIGO}{Advanced LIGO}
\newacro{AEI}{Albert Einstein Institute}
\newacro{CBC}{Compact Binary Coalescence}
\newacro{CDS}{Control and Data System}
\newacro{CIT}{Caltech}
\newacro{E@H}{Einstein at Home}
\newacro{EM}{Electro-magnetic}
\newacro{ER}{Engineering Run}
\newacro{GCN}{Gamma-ray Coordinates Network}
\newacro{GLUE}{Grid LSC User Environment}
\newacro{GraCEDb}{Gravitational-wave Candidate Event Database}
\newacro{GRINCH}{Gravitational-wave Candidate Event Handlers}
\newacro{GSL}{GNU Scientific Library}
\newacro{HTC}{High Throughput Computing}
\newacro{IFO}{Interferometer}
\newacro{ICTS}{International Centre for Theoretical Sciences}
\newacro{IdP}{Identity Provider}
\newacro{iLIGO}{Initial LIGO}
\newacro{IUCAA}{Inter-University Centre for Astronomy and Astrophysics}
\newacro{KDC}{Key Distribution Center}
\newacro{LALSuite}{LSC Algorithm Libraries}
\newacro{LIAM}{LIGO Identity and Access Management}
\newacro{LDR}{LIGO Data Replicator}
\newacro{LDG}{LIGO Data Grid}
\newacro{LIGOdv}{LIGO Data Viewer}
\newacro{LIGOdv-web}{LIGO Data Viewer Web Service}
\newacro{LOSC}{LIGO Open Science Center}
\newacro{LSC}{LIGO Scientific Collaboration}
\newacro{LSCSoft}{LSC Software Repositories}
\newacro{LSST}{Large Synoptic Survey Telescope}
\newacro{LVAlert}{LIGO-Virgo Alert System}
\newacro{LVC}{\ac{LSC} and the Virgo Collaboration}
\newacro{LVCN}{LIGO Virgo Computing Network}
\newacro{KAGRA}{Kamioka Gravitational Wave Detector}
\newacro{NDS}{Network Data Server}
\newacro{ODC}{Online Detector Characterization}
\newacro{OSG}{Open Science Grid}
\newacro{PE}{Parameter Estimation} 
\newacro{RDS}{Reduced Data Set}
\newacro{SP}{Service Provider}
\newacro{TACC}{Texas Advanced Computing Center}
\newacro{XSEDE}{Extreme Science and Engineering Discovery Environment}
\newacro{VDT}{Virtual Data Toolkit}
\newacro{EO}{Engineering and Operations}
\newacro{COS}{Collaboration Operations Support}
\newacro{ABB}{Application Building Blocks}
\newacro{DHS}{Data Handling Services}
\newacro{DetChar}{Detector Characterization}
\newacro{GDS}{Global Diagnostics System}
\newacro{DA}{Data Analysis}
\newacro{MOU}{Memorandum of Understanding}
\newacro{DetChar}{Detector Characterization}
\newacro{VALID}[VALID]{Vetting Automation and Literature Informed Database}

\newacro{H1}{LIGO Hanford}
\newacro{L1}{LIGO Livingston}
\newacro{V1}{Virgo}
\newacro{FAST}{Five-hundred-meter Aperture Spherical radio Telescope}
\newacro{LIGO}{Laser Interferometer Gravitational-Wave Observatory}
\newacro{ASKAP}{Australian Square Kilometre Array Pathfinder}
\newacro{CHIME}[CHIME]{the Canadian Hydrogen Intensity Mapping Experiment}
\newacro{GCN}[GCN]{Gamma-ray Coordinates Network}
\newacro{Fermi}{Fermi Gamma-ray Burst Monitor}
\newacro{SWIFT}{The Neil Gehrels Swift Observatory}
\newacro{O1}[O1]{the first observing run of Advanced LIGO and Advanced Virgo}
\newacro{O2}[O2]{the second observing run of Advanced LIGO and Advanced Virgo}
\newacro{O3}[O3]{the third observing run of Advanced LIGO and Advanced Virgo}
\newacro{O3a}[O3a]{the first part of the third observing run of Advanced LIGO and Advanced Virgo}
\newacro{O3b}[O3b]{the second part of the third observing run of Advanced LIGO and Advanced Virgo}
\newacro{O4}[O4]{the fourth observing run of Advanced LIGO, Advanced Virgo and Kagra}
\newacro{IPN}{Interplanetary Gamma-Ray Burst Timing Network}

\newacro{GW}{gravitational wave}
\newacro{GRB}{gamma-ray burst}
\newacro{lGRB}[lGRB]{long-duration gamma-ray burst}
\newacro{sGRB}[sGRB]{short-duration gamma-ray burst}
\newacro{FRB}{fast radio burst}
\newacro{KNe}{Kilonovae}
\newacro{KN}{Kilonova}
\newacro{BBH}{binary black hole}
\newacro{EM}{electromagnetic}

\newacro{SNR}[SNR]{signal-to-noise ratio}
\newacro{RA}[RA]{right ascension}
\newacro{DEC}[Dec]{declination}
\newacro{ADI}[ADI]{accretion disk instability}
\newacro{BH}[BH]{black hole}
\newacro{BNS}[BNS]{binary neutron star \acused{NS}}
\newacro{CSG}[CSG]{circular sine--Gaussian}
\newacro{SG}[SG]{sine--Gaussian}
\newacro{GBM}[Fermi-GBM]{Fermi Gamma-ray Burst Monitor}

\newacro{GW}[GW]{gravitational wave}
\newacro{NS}[NS]{neutron star}
\newacro{NSBH}[NSBH]{neutron star-black hole \acused{NS} \acused{BH}}

\newcommand{\sGRBrateApp}{$303_{-300}^{+1580}$}
\newcommand{\sGRBrateTopHatA}{$6.15_{-6.06}^{+31.2}$}
\newcommand{\sGRBrateTopHatB}{$3.34_{-3.29}^{+16.7}$}

\hypersetup{draft}

\title[Short gamma-ray burst rate]{The apparent and cosmic rates of short gamma-ray bursts}

\author[]{E. J. Howell$^{1}$\thanks{E-mail:
XXX@uwa.edu.au}, E. Burns$^2$ and A. Goldstein$^3$\\\\
$^1$ OzGrav-UWA, School of Physics and Astrophysics, University of Western Australia, Crawley WA 6009, Australia\\
$^2$ Department of Physics and Astronomy, Louisiana State University, Baton Rouge, LA 70803 USA\\
$^3$ Science and Technology Institute, Universities Space Research Association, Huntsville, AL 35805, USA\\
}

\begin{document}


\maketitle
\begin{abstract}
The short gamma-ray burst (sGRB), GRB~170817A, is often considered a rare event. However, its inferred event rate, $\mathcal{O}(100s)\ \text{Gpc}^{-3}\ \text{yr}^{-1}$, exceeds cosmic sGRB rate estimates from high-redshift samples by an order of magnitude. This discrepancy can be explained by geometric effects related to the structure of the relativistic jet. We first illustrate how adopting a detector flux threshold point estimate rather than an efficiency function, can lead to a large variation in rate estimates. Simulating the Fermi-GBM sGRB detection efficiency, we then show that for a given a universal structured jet profile, one can model a geometric bias with redshift. Assuming different jet profiles, we show a geometrically scaled rate of GRB~170817A is consistent with the cosmic beaming uncorrected rate estimates of short $\gamma$-ray bursts (sGRBs) and that geometry can boost observational rates within $\mathcal{O}(100s)$\,Mpc. We find an apparent GRB~170817A rate of $303_{-300}^{+1580}$ $\mathrm{Gpc}^{-3}\, \mathrm{yr}^{-1} $ which when corrected for geometry yields $6.15_{-6.06}^{+31.2}$ $\mathrm{Gpc}^{-3}\, \mathrm{yr}^{-1} $ and $3.34_{-3.29}^{+16.7}$ $\mathrm{Gpc}^{-3}\, \mathrm{yr}^{-1} $ for two different jet profiles, consistent with pre-2017 estimates of the isotropic sGRB rate. Our study shows how jet structure can impact rate estimations and could allow one to test structured jet profiles. We finally show that modelling the maximum structured jet viewing angle with redshift can transform a cosmic beaming uncorrected rate to a representative estimate of the binary neutron star merger rate. We suggest this framework can be used to demonstrate parity with merger rates or to yield estimates of the successful jet fraction of sGRBs.
\end{abstract}

\begin{keywords}
gamma-rays: bursts -- gamma-ray: observations -- methods: data analysis -- cosmology: miscellaneous
\end{keywords}

\graphicspath{{./figures/}}

\section{Introduction}
Short gamma-ray bursts (sGRBs) represent one of the most energetic phenomena in the universe, arising from the merging of compact binary systems such as neutron stars. These events are pivotal for advancing our knowledge in high-energy astrophysics, gravitational waves, and the behavior of matter under extreme conditions. The landmark observation of GRB~170817A \citep{Goldstein_2017ApJ,savchenko_integral_2017}, coupled with the detection of gravitational waves from the binary neutron star merger (BNS) GW170817 \citep{LSC_BNS_2017PhRvL,LSC_GW_GRB_2017ApJ} prompted an electromagnetic followup campaign that enriched our understanding of the electromagnetic counterparts to gravitational-wave events \citep{LSC_MM_2017ApJ,resmi_low_2018,margutti_binary_2018}. However, it also uncovered significant discrepancies between the estimated sGRB rates from models predating 2017 and those derived from the observation of GRB~170817A.

Table \ref{table:rate_estimates} highlights a number of recent studies in which a lower limit on the sGRB event rate has been estimated using the distance of GRB~170817A; the term \emph{lower limit} is used here as these numbers are assumed uncorrected for beaming effects. The table provides short notes on the methodologies employed and extrapolates the implied maximum distance unless defined in the study and shows that the median 170817A-like rates range from 190 -- 370 \rate. These numbers are much higher than pre-170817 estimates based on the more distant sGRB sample with available redshifts which were, at most, $\mathcal{O}(10)$ \rate \citep{Nakar_2006ApJ, coward_swift_2012, WandermanPiran2015MNRAS, Mandel2022LRR}. \\
The discrepancy between the rates can be reconciled through consideration of the structured jet of GRB 170817A \citep{Salafia_2016MNRAS,Salafia_Ghirlanda_2022,Salafia2023A&A}. A structured jet  profile, in which an ultra-relativistic core smoothly transforms to a milder relativistic outflow at greater angles, was predicted more than a decade previous to GRB 170817A \citep{lipunov_gamma-ray_2001,rossi_afterglow_2002,zhang_gamma-ray_2002}. This possibility was an explanation of the low luminosity of GRB 170817A even though it was observed at a distance of 42\,Mpc. It was finally confirmed with late electromagnetic observations that proved the emergence of an off-axis jet \citep{lazzati_late_2017,Mooley_2018,Alexander_2018ApJ}. It is clear that the low observed flux from a wide viewing angle, coupled with the relatively close proximity, where components of a wider jet become detectable, can cause an elevation in the event rate through geometric factors. 

How to reconcile the discrepancies between the geometrically biased rate estimates of locally observed bursts and the estimates from higher-$z$ events is the focus of this study. We will explore if a
refined model that incorporates detection efficiencies and structured jet profiles can reconcile the apparent discrepancies in low-$z$ and high-$z$ event rates. We will also highlight how a more detailed description of detector efficiency can result in improved rate estimates.

The structure of the paper is organised as follows. In Section \ref{section_rate_Framework} we introduce our framework to estimate sGRB rate densities. We compare two methodologies: one employing a simple point estimate and another utilizing a detector efficiency function. Section \ref{section_GBM_efficiency} presents the detection efficiency function for the Fermi Gamma-ray Burst Monitor (GBM), and Section \ref{section_GBM_eff_with_z} calculates the efficiency as a function of redshift. In Section \ref{section_impact_det_sens_on_rates}, we explore the impact of two different rate estimation methods and demonstrate the advantages of incorporating detection efficiencies. Section \ref{section:sj_models} details two structured jet models employed in our study, outlining their theoretical foundations and relevance to our analysis. Section \ref{section:effect_of_jet_geometry_on_rates} examines the effect of jet geometry on event rates and introduces a geometric relationship between apparent low-$z$ and high-$z$ sGRB rates. In Section \ref{section:sGRB_rate_density}, we derive an apparent low-$z$ sGRB event rate and demonstrate how a geometric scaling function can provide parity between low-$z$ and high-$z$ estimated sGRB rates. Section \ref{Sect:bns_rates_from_sgrbs} extends our framework to compare sGRB rates with BNS rates to confirm parity or to estimate the fraction of successful sGRB jets. We conclude in Section \ref{section:conclusions}, discussing the impact of our findings on our understanding of cosmic sGRB event rates, highlighting potential areas for future research.

Throughout this paper we assume a `flat-$\Lambda$' cosmology with cosmological parameters $\Omega_{\mathrm{m}}=0.315$ and $H_{0}=67.3$ km s$^{-1}$ Mpc$^{-1}$ \citep{Planck2020A&A}. Unless otherwise stated, peak flux values are assumed in the 64~ms time-window and the 50--300\,keV energy range. We present our main results using the maximum a posteriori estimates paired with asymmetric credible intervals. This approach facilitates direct comparisons with estimates from other studies obtained using non-Bayesian frameworks.

\begin{table*}
\begin{center}
\begin{tabular}{lccl}
  \hline
  Study                               & $d_{\mathrm{L,max}}$  & $\Rate_{\mathrm{170817A}}$ & \hspace{40mm}notes \\
                                      & (Mpc)                 & (\rate)                    &                     \\
  \hline
  \citet{ZhangBB2018NatCo}$\dagger$   & 65                    &  $190^{+440}_{-160}$ &  Defined as a lower limit estimate of sGRBs - $\Rate (L_{\mathrm{iso}} > 1.6 \times 10^{47}$ \lum \\
 \citet{DellaValle_2018MNRAS} $\ddagger$ & 49 & $352^{+810}_{-281}$ & The 170817A-like rate represents an observed rate of sGRBs associated with kilonovae.\\
 \citet{Salafia2022} $\dagger\dagger$ & -   &  $342^{+1798}_{-337}$ & The sensitive volume is integrated to $\infty$ so no $d_{\mathrm{L,max}}$ employed.\\
 \citet{Burgess2020FrASS} $\sharp$ & 4.8 & 170   & Low value of $d_{\mathrm{L,max}}$ due to low assumed sensitivity limit. \\
 \hline  
  \end{tabular}
    \caption{A selection of studies in which a lower limit on the sGRB event rate $\Rate$ has been estimated based on GRB~170817A. We state a maximum distance, $d_{\mathrm{L,max}}$, for cases in which that value has been used in the calculation (i.e. as an upper limit in equation \ref{equation_VT1}).
 $\dagger$ The value of $d_{\mathrm{L,max}}$ determined by simulating GRB~170817A light curves at progressively greater distances 
    $\ddagger$ Estimated using a $d_{\mathrm{L,max}}$ based on a Fermi-GBM 1.024\,s flux threshold $P_{\mathrm{L}}=0.5$ \pflux in the  50-300 kev band.
  $\dagger\dagger$ Calculated through eq (\ref{equation_VT2}) using a derived Fermi-GBM efficiency function for 170817A-like events estimated from data in the 10-1000 keV reporting band. 
  $\sharp$ A simple approximation based on a flux limit for Fermi-GBM of $10^{-7}$ \eflux $\sim$ 0.5 \pflux.  
    }
    \label{table:rate_estimates}
  \end{center}
\end{table*}


\section{The framework to calculate the rate density of sGRBs}
\label{section_rate_Framework}

Given a volumetric rate of some transient astrophysical population, $\Rate$, the mean number of events is given by $\Rate \langle V T \rangle$ where, for a given population, this is the product of the observation time $T$ and the volumetric reach of the search $V$. The angle brackets denote that we consider the rate to be constant or non-evolving in the co-moving frame; this assumption is supported as rate estimates will be dominated by the closest detected events. 

The posterior probability distribution of the event rate $\Rate_{i}$ for an individual detection $i$ is given by:

\begin{equation}
P(\Rate_{i}|1) \propto
 P(1|\Rate_{i})\,P(\Rate_{i})\,.
\label{eq_rate_posterior}
\end{equation}

Here, the first term is the likelihood function taken as the Poisson probability of observing one event given a mean rate $\Rate_{i} \langle V_{i} T_{i} \rangle $:

\begin{equation}
P(1|\Rate_{i}) =  \Rate_{i} \langle V_{i} T_{i} \rangle\, \exp(-\Rate_{i} \langle V_{i} T_{i} \rangle)\,,
\label{eq_rate_likelihood} 
\end{equation}

The second term of Equation~\ref{eq_rate_posterior} is the prior; for estimates of $\Rate_{i}$ that typically span several orders of magnitude, a good choice is the Jeffreys prior $\Rate_{i}^{-1/2}$, which allows an equal probability per decade. In this case, the posterior distribution becomes:

\begin{equation}
P(\Rate_{i}|1) \propto  \Rate_{i}^{1/2}\, \langle V_{i} T_{i} \rangle \, \exp(-\Rate_{i} \langle V_{i} T_{i} \rangle)\,.
\label{eq_rate_posterior_final}
\end{equation}

To estimate the sensitive time-volume $\langle V T \rangle$, two approaches are generally used in the literature:

\subsection{Method [A]: Using a detector sensitivity threshold point estimate} 
\label{sect_estimating_vt_method_A}
Given a single point estimate of the threshold sensitivity threshold of a detector, one can calculate the following integral:

\begin{equation}
\label{equation_VT1}
\langle VT \rangle= 
T \int^{z_{\mathrm{M}}}_{0} \frac{\mathrm{d}V_{\mathrm{c}} (z) }{\mathrm{d}z } \frac{ 1 }{( 1 + z)}\,\mathrm{d}z\
\end{equation}

\noindent where $T$ is the total observation time and the factor $1/(1 + z)$ factor corrects clock rates from the source frame to the observer frame. The upper limit of the integral is set by the maximum detection redshift of a given burst, $z_{\mathrm{M}}$ in terms of its threshold sensitivity \citep{Soderberg2006Natur, coward_swift_2012,howell_2013,DellaValle_2018MNRAS}. The co-moving cosmological volume element $\mathrm{d} V_{\mathrm{c}} (z) /\mathrm{d}z$  describes how the number densities of non-evolving objects locked into Hubble flow are constant with redshift and is given by:

\vspace{0mm}
\begin{equation}\label{dvdz}
\frac{ \mathrm{d} V_{\mathrm{c}}}{\mathrm{d}z}= \frac{4\pi c}{H_{0}}\frac{D_{\mathrm{L}}^{2}(z)}{(1 +
z)^{2}E(z)}\,,
\end{equation}

\noindent with $D_{\mathrm{L}}$ is the luminosity distance and $E(z)$ the normalized Hubble parameter,
\vspace{-0.5mm}
\begin{equation}\label{hz}
E(z)\equiv H(z)/H_0 = \big[\Omega_{\mathrm m} (1+z)^3+ \Omega_{\mathrm \Lambda} \big]^{1/2}\,.
\end{equation}
%

The threshold sensitivity required to define $z_{\text{M}}$ is obtained though flux threshold point estimate, $P_{\mathrm{L}}$, for astrophysical populations of GRBs. Therefore, if adopting equation (\ref{equation_VT1}), the maximum detection redshift of a GRB, $z_{\mathrm{M}}$, can be determined numerically by solving:

\begin{equation}\label{eq_max_distance}
D_{\mathrm{L}}(z_{\mathrm{M}})^{2} =   \frac{L}{4 \pi P_{\mathrm{L}}} \frac{C_{\text{N}}^{-1}}{ k(z_{\mathrm{M}})}\frac{1}{1 + z}\,.
\end{equation}

\noindent where, $L$ is the isotropic equivalent luminosity (erg s$^{-1}$) in the 1--$10^{4}$ keV rest-frame band and $P_{\mathrm{L}}$ is a photon flux (ph s$^{-1}$\,cm$^{-2}$). The $(1 + z)$ factor in the denominator is included as the standard definition of $D_{\mathrm{L}}(z)$ is valid for an energy flux, but we convert to photon counts \citep{meszaros_cosmological_2011}. Equation (\ref{eq_max_distance}) includes two corrections that assume a given $\gamma-$ray spectrum photon $N(E)$ (in ph\,s$^{-1}$ cm$^{-2}$ keV$^{-1}$).

The normalization constant \(C_{\text{N}}\) calculates the average energy per detected photon within the specified energy range $\left[E_{1 }, E_{2 }\right]$:

\begin{equation}\label{eq_bol_correction}
C_{\text{N}} = \frac{\int^{10\,\text{MeV}}_{1\,\text{keV}} E N(E) \, \mathrm{d}E}{\int^{\text{E}_{2}}_{\text{E}_{1}} N(E) \, \mathrm{d}E}\,.
\end{equation}

\noindent This constant has units of ergs and normalizes the observed gamma-ray data to a standard measurement that would be recorded for a burst hypothetically observed at $z=0$.
The function, $k(z)$ is the cosmological $k$-correction for the given spectrum at redshift $z$,

\begin{equation}\label{eq_bol_correction}
k(z)=\frac{\int_{E_{1 }}^{E_{2 }} N(E) \mathrm{d} E}{\int_{(1+z) E_{1 }}^{\left(1+z E_{2}\right.} N(E) \mathrm{d} E},
\end{equation}

\noindent which compensates for the effect of cosmic downshifting of the spectrum relative to a fixed detector bandwidth. 

%
%
%
%
%
%

\begin{figure}
 \centering
 
 \includegraphics[scale = 0.60,origin=rl]{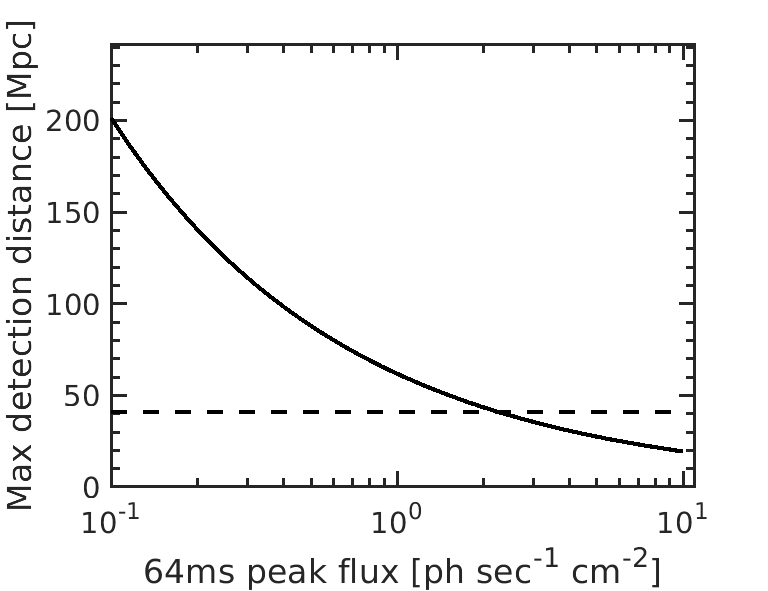}
 \caption{The maximum distance at which 170817A would have been detected is plotted against a range of GBM flux sensitivity limits. This plot assumes an apparent luminosity for a burst observed outside the jet axis; therefore, it is not the true maximum distance based on the intrinsic properties of the burst. The dashed horizontal line indicates the distance at which GRB~170817A was detected corresponding to a peak photon flux of around 2.27 \pflux.}
  \label{fig:fig_max_dist_with_flux_lim}
\end{figure}

Figure \ref{fig:fig_max_dist_with_flux_lim} illustrates the quantity, $d_{\mathrm{L}}(z_{\mathrm{M}})$, for GRB~170817A as a function of different peak flux sensitivity thresholds, $P_{\mathrm{L}}$. This curve is calculated for illustration using equation (\ref{eq_max_distance}) and the rest-frame luminosity and spectral parameters for the brightest part of GRB~170817A, modeled by a Comptonized spectrum with power law index $\alpha = -0.85$ and E$_{\mathrm{peak}}$ = 229\,keV \citep{Goldstein_2017ApJ} . Given the 64~ms peak flux of $2.3 \pm 0.8$ \pflux\, in the 50--300 keV GBM triggering band, by rearranging equation (\ref{eq_max_distance}) one can calculate the isotropic equivalent luminosity in the 1--$10^{4}$~keV rest-frame band of $(1.4 \pm 0.5 ) \times 10^{47}$ \lum.

Assuming a nominal detection threshold of 1 \pflux \, we see that this equates to a maximum detection distance of 61.8 Mpc ($z_{\mathrm{M}}$=0.014) for this event. Selecting a higher point estimate, 2 \pflux , results in a maximum detection distance of 43.5 Mpc ($z_{\mathrm{M}}$=0.0097). Based on the analysis in later section \ref{section_impact_det_sens_on_rates}, we find the difference in the relative quantities of $1/\langle VT \rangle $ calculated by adoption of the former or latter flux thresholds equates to a range of 140-470 \rate; a factor of over 3 in rate estimates. Thus significant variation in rates can occur through choices of a flux threshold point estimate. 

For GRBs, adopting the most realistic point estimate $P_{\mathrm{L}}$ is difficult as values vary considerably in the literature; this is shown later in Table \ref{table:flux_thresholds}. It is worth noting that initial point estimates from post-launch instrument specifications often differ from those based on statistical analyses of accumulated data once the instruments are operational. Additionally, such limits are often absolute thresholds only appropriate for very low efficiency. For example pre-launch scientific performance requirements for Fermi-GBM \citep[i.e.][]{Kippen2001AIPC} suggested a threshold of $<0.5$ \pflux in the 50--300keV triggering range: the actual detector efficiency at this level would be $<$0.1\% (see later Table \ref{table:flux_thresholds}). The importance of estimating detector sensitivity is obviously important, but not a trivial task without access to simulated detector efficiency curves \citep{Howell2014MNRAS}.


\subsection{Method [B]: Using a detector efficiency function} 
\label{sect_estimating_vt_method_B}
If detector efficiency curves are available one can calculate a detection efficiency with redshift, $\Sigma_{\mathrm{z}}(z)$ for a given source population \citep{Howell2014MNRAS}. One can fold this function with $\mathrm{d}V_{c}/\mathrm{d} z$ and calculate the integral of equation (\ref{equation_VT1}) out to $\infty$. 

\begin{equation}
\label{equation_VT2}
\langle VT \rangle= T
 \hspace{-1mm} 
 \int^{\infty}_{0} \frac{\mathrm{d}V_{\mathrm{c}} (z) }{\mathrm{d}z } \frac{ \Sigma_{\mathrm{z}}(z) }{( 1 + z)}\,\mathrm{d}z \,.
\end{equation}
\vspace{1.0mm}

\noindent See for example \citet{BNS_rate_supp_2016ApJS,Salafia2022} who have applied this approach to GWs and sGRBs respectively. Other than avoiding the dependence on a highly uncertain point estimate of flux sensitivity, this approach allows the integral $\langle VT \rangle$ to smoothly converge towards zero rather than enforcing an abrupt cutoff in $\langle VT \rangle$.

In the following section we will compare approaches to estimate event rates using both flux threshold point estimates and detection efficiency functions. We will first estimate a detection efficiency function for Fermi-GBM and the corresponding function 
$\Sigma_{\mathrm{z}}(z)$.

\section{GBM sGRB Peak Flux Detection Efficiency}
\label{section_GBM_efficiency}
To calculate the time-volume integral, eq. (\ref{equation_VT2}), an estimate of the peak flux detection efficiency function with redshift $\Sigma_{\mathrm{z}}(z)$ is required. The first stage in determining this component is to estimate the detection efficiency in peak flux space $\Sigma_{\mathrm{P}}(P)$.

To obtain the function $\Sigma_{\mathrm{P}}(P)$, there are a number of different factors to consider, such as the energy detection bandwidth or the use of different count rate thresholds above background. Furthermore, harder bursts typically have lower photon count rates than the softer long duration GRBs which can decrease the detection efficiency. One approach, as used by \citet{Salafia2022}, is to fit to the available Fermi-GBM data in the required band (the 10-1000 keV reporting band in that study) assuming the expected $P^{-3/2}$ integrated flux distribution.  The approach used here is to model the detector response given a simulated population of sGRBs \citep[see also][]{Lien_2014ApJ,Howell2014MNRAS,Shahmoradi2011MNRAS}.

We determine $\Sigma_{\mathrm{P}}(P)$ for the Fermi-GBM by recording the detection fraction of 64ms photon fluxes in the 50-300 keV band from a simulated population of sGRBs, where each GRB has a spectrum sampled from the GBM spectral catalog~\citep{Poolakkil21}. The simulated population of GRBs includes a variety of arrival angles, a variety of Fermi rocking angles and orbital locations sampled from historical data; this ensures that the model is representative of GBM's actual in-orbit performance. The simulations assume that at least 2 detectors can trigger a burst at $\geq 4.5$ sigma. Furthermore, the background rates and background variability observed over the first 13 years of GBM operation including entering and exiting the South Atlantic Anomaly are incorporated. We note that these simulations have not considered the FSW trigger algorithms but they represent a first-order approximation for the trigger efficiency in the 64\,ms timescale.

\begin{figure}
 \centering
 \includegraphics[scale = 0.60,origin=rl]{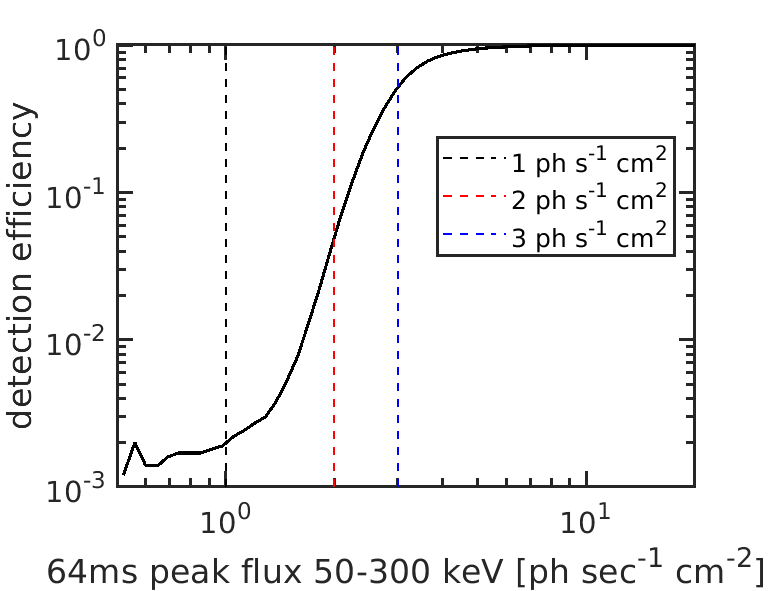}
 \caption{The Fermi-GBM detection efficiency $\Sigma_{\mathrm{P}}(P)$ for the 64-ms photon flux in the 50--300 keV band. The plot shows a rapid decline in sensitivity between 1 - 4 \pflux. Three fiducial flux limits are also indicated; the latter corresponds with the peak of the differential distribution of detected sGRBs shown in Figure \ref{fig:gbm_sgrb_flux_count}. The small feature at the lowest peak fluxes is an artifact due to the number of simulations used to create the efficiency curve.}
  \label{fig:GBM_Eff_in_50_300keV}
\end{figure}

\begin{figure}
 \centering
 \includegraphics[scale = 0.60,origin=rl]{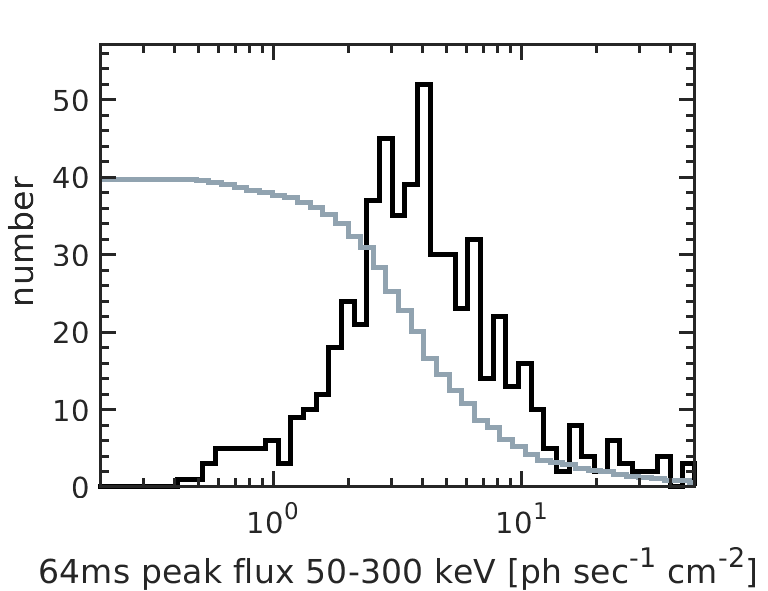}
 \caption{The Fermi-GBM detection distribution for the 64-ms photon flux in the 50--300 keV band (dark line). The peak of the distribution is around 3 \pflux, below which detection efficiency is less than 34\%. A small excess is evident at around 0.5-1 \pflux which is a result of non-uniform detector pointings. The cumulative distribution (grey line) is the 1-year averaged count.}
  \label{fig:gbm_sgrb_flux_count}
\end{figure}

Figure \ref{fig:GBM_Eff_in_50_300keV} shows the estimated Fermi-GBM efficiency curve for the 64-ms photon flux. The plot illustrates how the efficiency degrades rapidly below around 5 \pflux. At a flux limit of around 1 \pflux, above which around 95\% of sGRBs are detected the efficiency is of order $2 \times 10^{-3}$; at 2 \pflux the detection efficiency is still less than 1\% (0.5\%). We find that within the range 2 -- 4 \pflux the efficiency can be approximated (norm of residuals = 0.02) by the cubic function:

\begin{equation}
\Sigma_{\mathrm{P}} = \eta 1\,P^{3} + \eta 2\,P^{2} +\eta 3\,P + \eta 4\,,
\end{equation}

 \noindent with $\eta1$ = -0.14, $\eta2$ = 1.20, $\eta3$ = -2.87, $\eta4$ = 2.10.

Figure \ref{fig:gbm_sgrb_flux_count} shows the 64-ms peak flux distribution in the 50--300\,keV energy range for all bursts recorded with a T$_{90}$ less than 2.1\,s; this is total of 578 bursts from the Fermi-GBM sample
of 3447 bursts up to February 2023 \citep[14.5 years of data from:][]{fermigbrst2024}
The plot shows that the dominant peak of the distribution is around 3 \pflux, below which the detection losses are expected to be significant given an expected -5/2 power-law differential distribution. The efficiency at 3 \pflux is around 52\%, degrading significantly below this value as discussed above.
Fig \ref{fig:gbm_sgrb_flux_count} also highlights a small enhancement at around 0.5 --1  \pflux due to non-uniform detector pointings. The cumulative count is also shown and is scaled by 14.5 to represent detections per year; this curve becomes asymptotic at around 1 \pflux. Given the nominal -3/2 scaling above this point, we note that a factor of 2 in peak flux sensitivity corresponds to around an increase factor of 2.8 times as many bursts; thus the motivation to improve detector sensitivity is clear.

\section{GBM sGRB Detection Efficiency with redshift}
\label{section_GBM_eff_with_z}
\begin{figure}
 \centering
 \includegraphics[scale = 0.60,origin=rl]{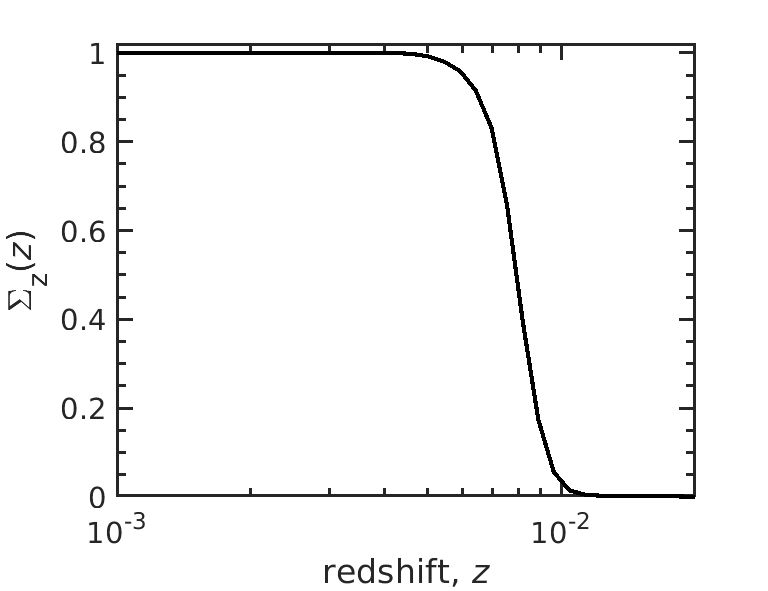}
 \caption{The Fermi-GBM detection efficiency with redshift to a GRB~170817A-type burst. }
  \label{fig:170817_eff_with_z}
\end{figure}

Once an estimate of the detector efficiency is obtained across a range of the chosen detection threshold parameter, i.e. peak flux, SNR, or magnitude, one can determine the detection efficiency through a range of redshift, $\Sigma_{\mathrm{z}}(z)$  \citep{Howell2014MNRAS,BNS_rate_supp_2016ApJS,Salafia2022}. One can estimate this function on a population level by sampling population parameters across a range of redshift bins. Alternatively, one can determine the function $\Sigma_{\mathrm{z}}(z)$ for a single representative source using best known intrinsic properties.

In this study we will use a small sample of Fermi-GBM sGRBs with known redshifts. By treating these events independently, we will construct $\langle VT \rangle$ for each and an associated rate. Due to its close proximity, we will find that the cosmic rate is dominated by GRB~170817A. Given the apparent intrinsic luminosity of GRB~170817A calculated in section \ref{sect_estimating_vt_method_A} one can calculate the observed peak flux at each step of $z$. The resulting efficiency $\Sigma_{\mathrm{z}}(z)$ is constructed by the mapping, $\Sigma_{\mathrm{z}}(z)$ = $\Sigma_{\mathrm{P}}(P_{z})$, where $P_{z}$ is the observed peak flux from a sGRB at given $z$. 

Figure \ref{fig:170817_eff_with_z} shows the resulting function $\Sigma_{\mathrm{z}}(z)$. We find that GRB~170817A would have been detectable by Fermi-GBM out to $z\sim$0.006 with 100\% efficiency. Interestingly, the detection efficiency for a GRB~170817A-like burst is only around 17\%. This relatively low efficiency may at first seem strange considering it was detected at 42 Mpc, but is purely a consequence of the viewing angle. The efficiency for this burst at more favorable orientations will be discussed later (see later in Figure \ref{fig:e_z_with_viewing_angle}).

An important point that must be emphasised here is that the function 
$\Sigma_{\mathrm{z}}(z)$ calculated above does not take into account geometrical factors that can impact detection efficiency. In particular, $\Sigma_{\mathrm{z}}(z)$ is that of a GRB~170817A-type burst, \emph{viewed at the same orientation with z}; in this case $\sim 20$\deg. As such the time-volume integral $\langle VT \rangle$ calculated as in equation \ref{equation_VT2} is an \emph{apparent} quantity and its use results in \emph{apparent} rates. This is an important consideration and will be discussed in detail in section \ref{section:effect_of_jet_geometry_on_rates}.

\section{The impact of detector sensitivity on event rate calculations}
\label{section_impact_det_sens_on_rates}
As described in the previous section, if detector efficiency curves are available one can construct and fold the function $\Sigma_{\mathrm{z}}(z)$ with $\mathrm{d}V_{c}/\mathrm{d} z$ and calculate the integral of equation (\ref{equation_VT2}) smoothly out to $z_{\mathrm{M}}$ without an abrupt cutoff.
In this section we will quantify the effect of using this approach to determine rate estimates (method [A] in section \ref{sect_estimating_vt_method_A} against using a point estimate (method [B] in section \ref{sect_estimating_vt_method_B}). 

\begin{figure}
 \centering
 \includegraphics[scale = 0.65,origin=rl]{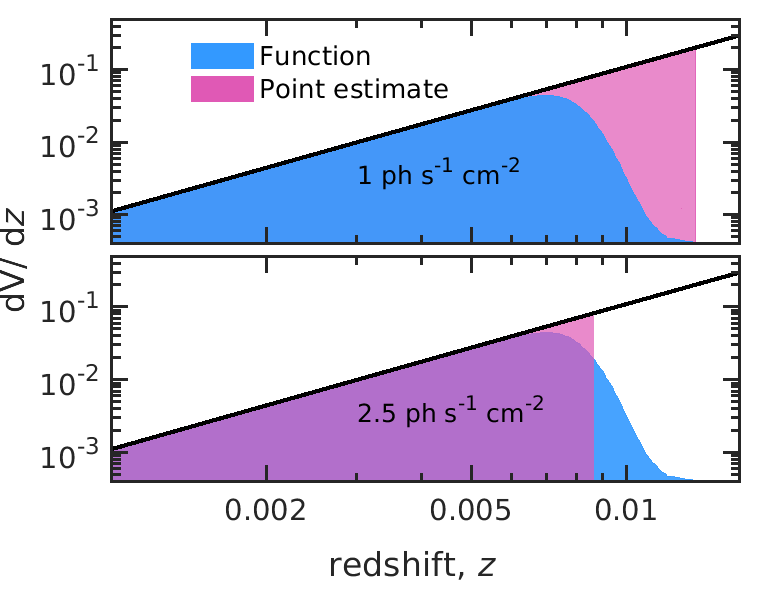}
 \caption{To illustrate difference in calculating equation $\langle V T \rangle$ using a flux efficiency function or a flux threshold point estimate, the relative areas of integration are shown for both cases. The top panel assumes a flux threshold of 1.0 \pflux -- in this case the integral would result in an overestimate in $V_{c}$ and thus, an underestimate in rate. The lower panel assumes a higher value point estimate 2.5 \pflux - this case will result in an overestimate of the event rate.}
  \label{fig:fig_volume_int}
\end{figure}

Figure \ref{fig:fig_volume_int} compares the regions of integration for the integrand of equation (\ref{equation_VT1}) based on two threshold point estimates $P_{\mathrm{T}} = [1,\,2.5]$ \pflux with the regions defined by equation (\ref{equation_VT2}) which includes a 
detection efficiency function $\Sigma_{\mathrm{z}}(z)$. In both panels the region of integration using the efficiency function begins to decline at around $z \sim 0.006$ - this would correspond with a value of around 3 \pflux at which differences in the detected $P^{-3/2}$ integrated rate distribution start to deviate significantly from the intrinsic population due to instrument sensitivity.

The top panel shows the case in which a flux threshold of 1.0 \pflux is applied; this produces an overestimate in $V$ in comparison to that determined using an efficiency function, and thus, an underestimate in rate, $\Rate$. The lower panel considers the opposing situation in which applying a higher flux threshold of 2.5 \pflux can result in a comparative underestimate of $\Rate$. 

\begin{figure}
 \centering
 \includegraphics[scale = 0.65,origin=rl]{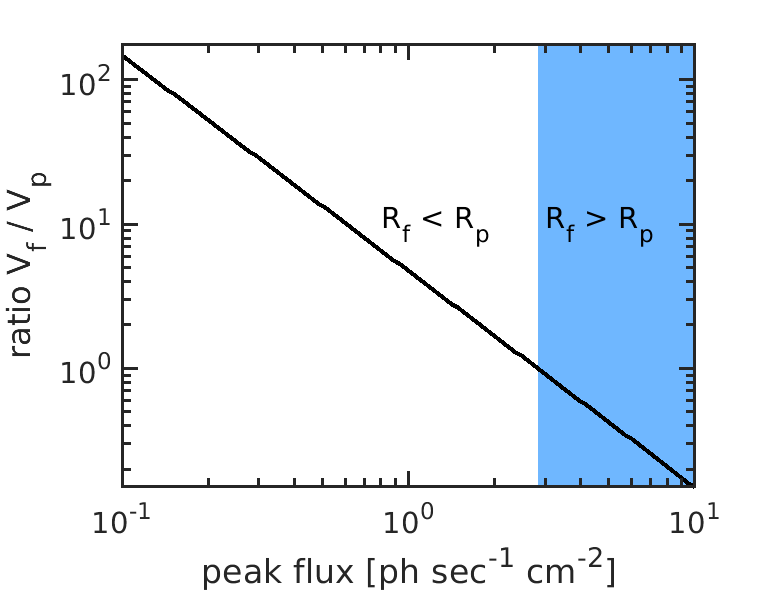}
 \caption{The ratio between the sensitive volumes calculated assuming a 
 flux efficiency function or a flux threshold point estimate for a range of flux detection thresholds. The plot shows that below around 2.8 \pflux use of a point estimate can overestimate the event rate.}
  \label{fig:fig_volume_ratio}
\end{figure}


Figure \ref{fig:fig_volume_ratio} quantifies the effective differences between the two approaches for the Fermi-GBM -- flux threshold point estimate (subscript p) against efficiency function (subscript f) -- for a range of flux thresholds. The effective error in $\langle VT \rangle$ is calculated through the ratio of the integrals:

\begin{equation}
\label{equation_volume_error}
\frac{\langle V \rangle_{\mathrm{f}}}{\langle V \rangle_{\mathrm{p}} }= \hspace{-1mm} 
 \int^{\infty}_{0} \frac{\mathrm{d}V_{\mathrm{c}} (z) }{\mathrm{d}z } \frac{ \Sigma_{\mathrm{z}}(z) }{( 1 + z)}\,\mathrm{d}z \Bigg/
 \hspace{-2mm} \int^{z_{\mathrm{M}}}_{0} \frac{\mathrm{d}V_{\mathrm{c}} (z) }{\mathrm{d}z } \frac{ 1 }{( 1 + z)}\,\mathrm{d}z\
\end{equation}
\vspace{1.0mm}

\noindent where $z_{\mathrm{M}}$ represents the maximum detection redshift and all other quantities are as for equations (\ref{equation_VT1}) and (\ref{equation_VT2}). The implications on rate estimates are apparent through the inverse ratio i.e. $\langle V \rangle_{\mathrm{f}}/\langle V \rangle_{\mathrm{p}}  \sim \Rate_{\mathrm{p}}/\Rate_{\mathrm{f}} $. One can see that up to around 2.8 \pflux \, a point estimate can lead to an overestimation in $\Rate$ (region indicated by no shading); above this point an underestimate in rate (light-blue shading). Examination of the lower panel of Fig. \ref{fig:fig_volume_int} at the intersection point 2.8 \pflux shows that the decreasing region of integration is approximately divided at this point.

We have noted a large range of Fermi-GBM flux thresholds provided in the literature that could be employed as point estimates. Within the 50-300 keV triggering band we find 64ms estimates ranging from $P_{\mathrm{L}}$ = 0.27 \pflux to $P_{\mathrm{L}}$ = 2.37 \pflux have been used in studies employing a range of approaches and assumptions; some estimate the rate, luminosity function and source rate evolution \citep[e.g.][]{WandermanPiran2015MNRAS,Zhang2018ApJ} while some have assumed a set form of the luminosity function, structured jet profile or binary neutron star merger rate estimates to calculate joint gravitational wave/sGRB rates \citep[e.g][]{Clark2015ApJ,Howell2019MNRAS,2022MNRAS_Patricelli}. The range of flux thresholds correspond to Fermi-GBM detector efficiencies in the range $\Sigma_{\mathrm{P}}(P)\sim 0.1 - 17.2 $ and thus $\langle V \rangle_{\mathrm{f}}/\langle V \rangle_{\mathrm{p}}$ values from 33.1 down to 1.3. 

It is clear that in calculations of rates using $1/\langle VT \rangle $, a choice of point estimate is highly sensitive can lead to a large variation in rate estimates. The use of a detector flux efficiency curve allows one to calculate the integral of equation (\ref{equation_VT2}) to $\infty$, negating the choice of a single flux limit. By folding in the detection efficiency $\Sigma_{\mathrm{z}}(z) $ will allow the integral to smoothly converge towards zero rather than have an abrupt cutoff and a related overestimate in $\langle VT \rangle$. Thus, given the availability of a detector efficiency function here, we will employ functions $\Sigma_{\mathrm{z}}(z)$ along with eq. (\ref{equation_VT2}) for estimating event rates.

It is worth noting in conclusion, this framework is equally applicable to other quantities referenced as detection thresholds; for example, taking a point estimate of a gravitational wave detection threshold in terms of a signal-noise-ratio and integrating out to the horizon distance without considering the corresponding function $\Sigma_{\mathrm{z}}(z)$ would result in a similar discrepancy.

\section{Apparent Structured Jet Luminosity}
\label{section:sj_models}
\subsection{Luminosity and the angular timescale}
A structured jet has an angular dependence on energy $\mathrm{d}E(\theta)/\mathrm{d}\Omega$ and bulk Lorentz factor $\Gamma(\theta)$. The structured jet profile can be described by an ultra-relativistic core without sharp edges that smoothly transforms to a milder relativistic outflow at greater angles \citep{lipunov_gamma-ray_2001,rossi_afterglow_2002,zhang_gamma-ray_2002, Salafia_2016MNRAS,Salafia_Ghirlanda_2022,Salafia2023A&A}. 

Using this formulation to also describe the angular dependence of luminosity, $\mathrm{d}L(\theta)/\mathrm{d}\Omega$, one can gain valuable insights on the form of the GRB luminosity function \citep{Salafia2023A&A} and allow inferences from an observed flux \citep{Howell2019MNRAS}. However, to apply this prescription requires an acknowledgment of certain limitations. In particular,the variability from internal shocks within the jet leads to a series of short emission pulses which can overlap and modify the emission \citep{Salafia2016MNRAS}. Since an energy is a cumulative measure that integrates all emissions throughout the burst, the impact of pulse overlaps on this quantity is diluted over the total burst duration. 

To understand the effect of pulse overlap, in what follows we will take a rather simplified approach. Firstly, one must consider the angular timescale, \( \tau_{\theta}= R_\gamma/\Gamma(\theta)^2\,c \), which represents the time difference between the arrival of photons emitted at 
zero latitude and at an angle $1/\Gamma$ at a $\gamma$-ray emitting radius, $R_\gamma$. A smaller value of \( t_{\text{ang}} \) indicates less overlap in the pulses.

To illustrate this, let us set the total observed GRB duration by $ \tau_{\text{GRB}} = 
\tau_{\text{E}} - \tau_{\text{J,B}} + \tau_{\theta} $ where $\tau_{\text{E}}$ is the activity timescale of the jet engine and $\tau_{\text{J,B}}$ is the timescale in which the jet breaks out from the merger ejecta. One can assume that $\tau_{\text{E}} \geq \tau_{\text{J,B}}$ to avoid most of jet energy to be deposited in the cocoon rather than the jet \citep{Ramirez-Ruiz2002MNRAS,Beniamini2020ApJ}. If we then assume $\tau_{\text{E}} > \tau_{\theta}$, the observed duration is set by $ \tau_{\text{GRB}} \sim \tau_{\text{E}}$ and an average isotripic equivelent luminosity can scale approximately with the energy $L_{\text{iso}} \sim E_{\text{iso}}/\tau_{\text{E}} \sim E_{\text{iso}}/\tau_{\text{GRB}}$ \citep{Salafia2020A&A,Salafia_Ghirlanda_2022}. Furthermore, if one additionally assumes that ratio of peak luminosity to the average luminosity has no angular dependence one can formulate a luminosity profile that takes the form of the jet profiles for energy  $\mathrm{d}E(\theta)/\mathrm{d}\Omega$. To continue we will therefore make the  important assumption that $\tau_{\text{E}} > \tau_{\theta}$ \citep{Salafia2020A&A}. We note however that this is dependent on the range of the parameters $\Gamma$ and $R_\gamma$ as well as the structured jet profile; we further discuss and illustrate this subtle effect in Appendix \ref{app:angular_time_scale} using a toy model.

\subsection{Structured jet models}
As the goal of this study is to investigate the effect of jet geometry on event rates we will employ two different structured jet profiles. We consider apparent forms of the jet profile which include the effects of relativistic beaming, which enhances the observed brightness at off-axis viewing angles.

The intrinsic structured jet profile describes the angular distribution of energy  $\mathrm{d}E(\theta)/\mathrm{d}(\Omega)$ and Lorentz factor  $\Gamma(\theta)$. This has been generally modelled using simple Gaussian or power-law jet profiles. Different to an intrinsic jet profile which is independent of the observer's perspective an apparent jet profile takes into account relativistic aberration effects that describes how the jet appears to an observer. 

We adopt here two models: the first is an intrinsic structure that takes a Gaussian form which is modulated by relativistic beaming effects to produce an apparent structure. The second is an analytic broken power law form derived to directly fit an apparent profile: 
\subsubsection{Apparent Gaussian profile}
This model (model A hereafter) has a Gausssian intrinsic structure defined by \citet{zhang_gamma-ray_2002}:

\begin{equation}
\label{eq:intrinsic_Gauss_profile}
\begin{aligned}
     &\frac{\text{d}L}{\text{d}\Omega}(\theta_{\mathrm{v}}) = L_{\mathrm{c}} \exp \left( -\frac{\theta_{\mathrm{v}}^{2}}{2\,\theta^{2}_{\mathrm{c}}} \right)\\
     &\Gamma (\theta_{\mathrm{v}}) = 1 + (\Gamma_{\mathrm{0}}-1) \exp \left( -\frac{\theta_{\mathrm{v}}^{2}}{2\,\theta^{2}_{\mathrm{c}}} \right)\\ 
\end{aligned}
\end{equation}

\noindent where $\text{d}L/\text{d}\Omega(\theta_{\mathrm{v}})$ is the luminosity per unit solid angle, and $\Gamma (\theta_{\mathrm{v}})$, the Lorentz factor of the emitting material. The parameter, $\theta_{\mathrm{v}}$, is the viewing angle, $\theta_{\mathrm{c}}$, a parameter that defines the sharpness of the angular profile. The parameters $L_{\mathrm{c}}$ and $\Gamma_{\mathrm{0}}$ are the maximal values of the core luminosity and Lorentz factor at the center of the jet.

The observed isotropic equivalent $\gamma$-ray luminosity, $4 \pi dL/d\Omega(\theta_{\mathrm{v}})$ for an observer positioned at an angle $\theta_{\mathrm{v}}$ from the axis of the jet is:

\begin{equation}
\label{eq:SJ_App_Lum_Profile}
    L(\theta_{\mathrm{v}})=\int^{\pi/2}_{0} \int^{2\pi}_{0}  \frac{\delta_{\text{D}}^{3}(\theta,\phi,\theta_{v})}{\Gamma(\theta)}\, \frac{dL}{d\Omega}(\theta)\, \mathrm{d}\phi\, \mathrm{sin}\,\theta\mathrm{d}\theta 
\end{equation}

\noindent Here, $\delta_{\text{D}}$ is the relativistic Doppler factor, cubed to account for time dilation and abberation effects; it is a function of the polar angle $\theta$, the azimuthal angle $\phi$ and the viewing angle $\theta_{\mathrm{v}}$. It takes the form:

\begin{equation}
\label{eq:doppler_factor}
    \delta_{\text{D}}(\theta,\phi,\theta_{v}) = \frac{1}{\Gamma(\theta)\, (1 - \beta(\theta) \,\mathrm{cos}\,\xi)}
\end{equation}

\noindent where $\beta(\theta)$ is the velocity of the outflow at an angle, $\mathrm{cos}\,\xi$, relative to the observer; this latter factor is given by.

\begin{equation}
\label{eq:doppler_cos_factor}
    \mathrm{cos}\,\xi (\theta,\phi,\theta_{v}) = \mathrm{cos}\, \theta\, \mathrm{cos}\, \theta_{\mathrm{v}} + \mathrm{sin}\, \theta \,\mathrm{sin}\, \phi\, \mathrm{sin}\, \theta_{\mathrm{v}}
\end{equation}
\subsubsection{Apparent broken power-law profile}

To circumvent strong theoretical constraints on the intrinsic profile, \citet{Salafia2023A&A} presented an analytical broken power law model of the apparent structured jet profile. This model (model B hereafter) takes the following form:

\begin{align*}
& L\left(\theta_{\mathrm{v}}\right)=L_{\mathrm{c}} \left[1+\left(\frac{\theta_{\mathrm{v}}}{\theta_{\mathrm{c}}}\right)^{4}\right]^{-\alpha_{\text{L}} / 4}\left[1+\left(\frac{\theta_{\mathrm{v}}}{\theta_{\mathrm{w}}}\right)^{4}\right]^{-\left(\beta_{\text{L}}-\alpha_{\text{L}}\right) / 4}  \\
\end{align*}

A full inference of universal observed sGRB properties yielded estimates for several characteristics, including the structure parameters, apparent luminosity function, and source evolution. We adopt the best parameters based on the flux-limited sample, which are: s=3.0, $\theta_{\mathrm{c}}$=3.0\deg, $\alpha_{\text{L}}$=4.9, $\beta_{\text{L}}$=1.9 and a truncation angle $\theta_{\mathrm{w}}$=64.5\deg.


\begin{figure}
 \centering
 \includegraphics[scale = 0.65,origin=rl]{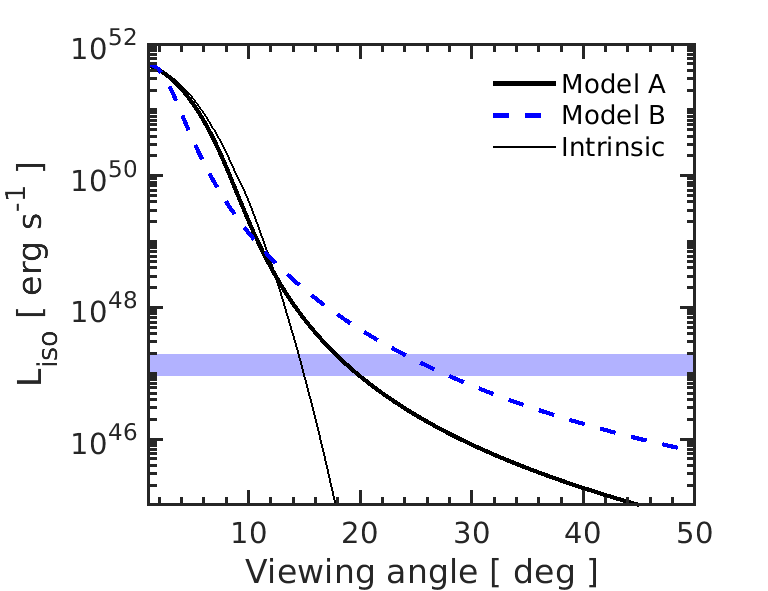}
 \caption{The apparent structured jet profiles adopted in this study are plotted as an isotropic equivalent luminosity based on a source at the distance of GRB~170817A . Model A is based on a Gaussian intrinsic profile (shown by the dashed curve) and model B a broken power law form based on a population inference of sGRBs \citep{Salafia2023A&A}. The shaded horizontal band shows the Fermi-GBM observed values of luminosity in the 50-300 keV band for GRB~170817A.}
  \label{fig:app_sj_profiles}
\end{figure}

\vspace{3mm}
Figure \ref{fig:app_sj_profiles} shows these two structured jet profiles assuming a source at the distance of GRB~170817A. The shaded horizontal band shows the two models are comparable with the observed isotropic equivalent luminosity of GRB~170817A, $1.4 \pm 0.5 \times 10^{47}$ \lum coinciding with a viewing angle of $20 \pm 5$ deg \citep{Mooley2018Natur}. The most likely maximum viewing angle maximum requires a consideration of the actual detection efficiency at a given flux. However, for an approximate estimate we note that in \citet{Goldstein_2017ApJ}
GRB~170817A would have been detected at 50\% of its observed brightness during an untargeted search and 44\% in a targeted search. Using a 50\% value of the observed flux in the 50-300keV band we find 1.1 \pflux which equates to an efficiency of
less than 1\%. We use this value to produce the blue horizontal band that infers that the maximum viewing
angle for a GRB~170817A-like burst would be 25 deg based on an isotropic
equivalent luminosity of  $7.1 \pm 2.4 \times 10^{46}$ \lum.


\section{Effect of jet geometry on rate estimates Rates}
\label{section:effect_of_jet_geometry_on_rates}

As widely understood, and as illustrated in Figure \ref{fig:fig_max_dist_with_flux_lim}, the detection of GRB~170817A would not have been possible had the burst originated from high-$z$ \citep[see also][]{Goldstein_2017ApJ}. Such a scenario would have negated the ability to observe the wider angled emissions of a structured jet that were accessible at lower-$z$. Understanding this allows us to make an important distinction that relates to the true meaning of Fig. \ref{fig:fig_max_dist_with_flux_lim}. The maximum distance in that plot represents that of a GRB~170817A-type burst, that is \emph{observed at the same viewing angle} with each increment of peak flux; in this case $\sim 20$\deg. Taking this interpretation, we immediately see that the \emph{true} maximum distance would correspond to the same burst observed face-on or within the core of the relativistic jet.

Although this realisation is quite intuitive, a framework for the relation between maximum viewing angle and distance relation and its implications was presented in \citet{Howell2019MNRAS}; based on the adopted structured jet profile and given Fermi-GBM flux limit, that study suggested a maximum viewing angle of around $\sim 22$\deg\, at the distance of GRB~170817A. Their Fig.\,5 suggested that the true (on-axes) maximum distance was order $z = 1$. We will revisit this relation in section \ref{sub_section_app_and_top_hat}.

The distinction between a low-$z$ sGRB observed through a wide-angled emission or a distant event observed on-axis has important implications for event rate estimates. For the low-$z$ scenario, the estimated $\langle VT \rangle$ can result in elevated estimations of the intrinsic event rate; the perceived rarity of the event, in comparison with the more general source population, is artificially boosted by geometrical effects. To illustrate this further it is useful to revisit the effect on the function $\Sigma_{\mathrm{z}}(z)$ in equation (\ref{equation_VT2}).

\begin{figure}
 \centering
 \includegraphics[scale = 0.65,origin=rl]{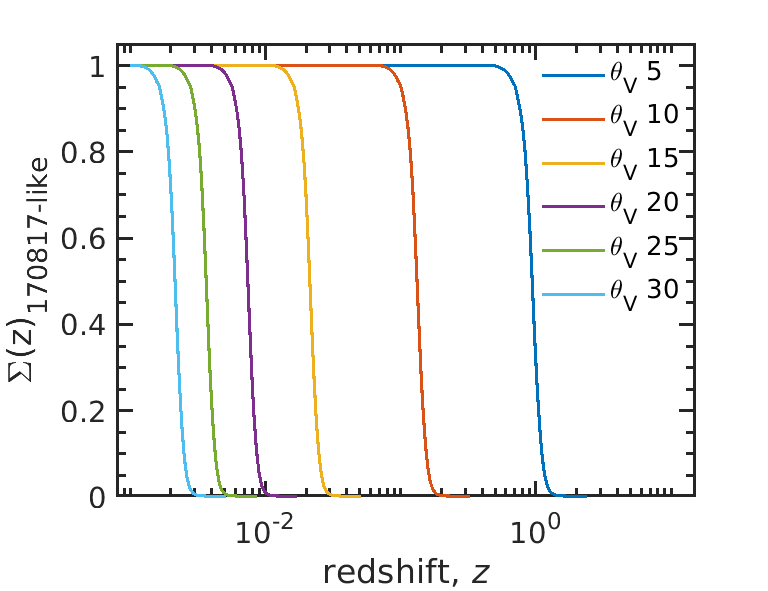}
 \caption{The Fermi-GBM detection efficiency with redshift for a GRB~170817A-type burst viewed at a range of viewing angles. In comparison with Fig \ref{fig:170817_eff_with_z} we see that the detection efficiency is optimised for more on-axis bursts that occur at higher-$z$ which due to source rate evolution are predominantly more frequent. }
  \label{fig:e_z_with_viewing_angle}
\end{figure}

Figure \ref{fig:e_z_with_viewing_angle} shows how, for a burst with a 170817-like structured jet profile (we assume model A), the function $\Sigma_{\mathrm{z}}(z)$ is dependent on the viewing angle. As shown by the ranges of the efficiency curves with $z$, observations at a wider viewing angle would have a low-flux contributions and thus only be accessible at low-$z$; in this case the $z$ contribution to the integral over $ \mathrm{d}V_{\mathrm{c}} (z)/\mathrm{d}z$ would be minimized resulting in a small $\langle VT \rangle$ and thus an inflated rate estimate. The inverse scenario corresponds to a burst observed on-axis within 5\deg. In this case the detection efficiency is optimal to nearly $z \sim 1$, so the integral limits over $z$ would result in a larger $\langle VT \rangle$ and a lower rate estimate. Its interesting to note that due to source rate evolution, the higher-$z$ bursts that are observed most frequently will contribute less to event rate estimates than the rarer low-$z$ observations.

\begin{figure}
 \centering
 \includegraphics[scale = 0.65,origin=rl]{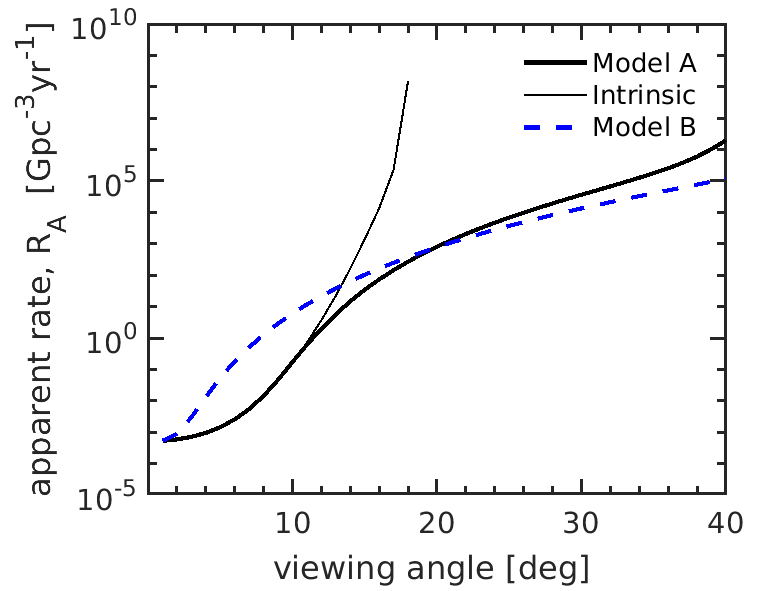}
 \caption{The apparent rate inferred by a GRB~170817A-like burst at 42\,Mpc observed at different viewing angles. The plot shows how two different models (as shown in Fig.\,\ref{fig:app_sj_profiles}) show apparent rate estimates that increase with viewing angle.}
  \label{fig:app_rate_with_viewing_angle}
\end{figure}

Figure \ref{fig:app_rate_with_viewing_angle} follows from Fig. \ref{fig:e_z_with_viewing_angle} to show how the inferred apparent rate will depend on the viewing angle of the sGRB. The rate is shown as a function of viewing angle for the 2 different structured jet models shown in Fig.\,\ref{fig:app_sj_profiles}. The plot shows how GRB~170817A, with an expected viewing angle of order 20\deg, would suggest an apparent rate $\mathcal{O}(100s)$\,\rate. The plot also shows consideration only of an intrinsic jet profile (for the case of a Gaussian profile) as opposed to an apparent profile would lead to vastly inflated rate estimates. This effect is due to the steep drop off in flux within 20\deg for the model presented in Fig.\ref{fig:app_sj_profiles}. 

Regardless of the model considered, Fig.\,\ref{fig:app_rate_with_viewing_angle} reinforces how larger viewing angles correspond to larger rate estimates. The plot also suggests how the field may evolve: we see that an improved understanding of the event rate, coupled with firm observational estimates of $z$ and and good estimates of viewing angle could eventually allow one to place constraints on structured jet profiles. Such a scenario may be possible by coupling higher sensitivity $\gamma$-ray detectors with multi-messenger observations.

The discussion in this section emphasizes how ignoring the effect of a source orientation can lead to possible false interpretations. For example, seemingly anomalous low-$z$ events of a particular class could be suggested as members of a under-luminous sub-population \citep[i.e.][]{howell_2013, Siellez2016arXiv}. Thus, such outliers could be a manifestation of geometry rather than time-volume effects \citep[see also the related discussion in][]{Salafia_2016MNRAS}. 

One should note that for high-$z$ GRBs, a \emph{top-hat} profile is a reasonable assumption, so a simple scaling can be applied between observed and intrinsic rates. At low-$z$ however the extended angular structure of the jet becomes more significant in influencing rate estimates. Understanding the relationship between detection and geometry for these lower-$z$ detections can provide useful insights into the actual rarity and rate of GRB~170817A-type bursts. To clarify the discussions that follow, it's important to define three different rates that can be inferred from observations:

%
%

\begin{description}
    \item [\bf{Apparent local rate, $\Rate_{\mathrm{A,L}}$}:] This is the rate inferred by observing a low-$z$ (GRB~170817A-like) burst for which the wider angled emissions of a structured jet are detectable. This is a geometrical dependent rate and for a given $z$ is a direct result of the jet profile of an individual burst. If one approximately assumes that on axis emissions would be from within $\sim10$\deg\,, Fig.\,\ref{fig:e_z_with_viewing_angle} suggests that this rate would dominate for bursts observed within $z \sim 0.05$. \\
    
    \item [\bf{Apparent cosmological rate, $\Rate_{\mathrm{A,C}}$}:] This is the inferred rate of sGRBs observed from outside a local volume $z \sim 0.05$ which are predominantly \emph{on-axis}. At such distances the sensitivity to the prompt emission will not be sufficient to detect the lower-flux contributions from the wings of a structured jet. This rate is also termed apparent, as at high-$z$, detection is dependent on the orientation of the relativistic beaming. \\
    
    \item[\bf{Intrinsic rate, $\Rate_{\mathrm{I}}$}:] This is the true intrinsic rate of occurrence in the Universe, regardless of beaming or related geometry. In this paper we will assume the majority of sGRBs result from BNS mergers, so can take the approximation $\Rate_{\mathrm{I}}\sim \Rate_{\mathrm{BNS}}$.
\end{description}

Table \ref{table:rate_parameters} presents the parameters employed in this study, including the above rate definitions, viewing angles, and detector efficiencies. These parameters are crucial for analyzing the observational data and theoretical models discussed herein.

\begin{table}
\begin{center}
\begin{tabular}{ll}
  \hline
   Parameter                     & Definition \\
  \hline
 $\Rate_{\mathrm{I}}$             &  Intrinsic rate of sGRB progenitors            \\
 $\Rate_{\mathrm{A,L}}$            & Apparent local low-$z$ rate of sGRBs             \\
 $\Rate_{\mathrm{A,C}}$           &  Apparent cosmological rate of sGRBs            \\
 $ \Rate_{\mathrm{BNS}}$          &  Intrinsic rate of binary neutron star mergers            \\           
 
  $P_{\mathrm{O}}$                  & Observed flux\\
  $P_{\mathrm{T}}$                  & Flux threshold point estimate\\
 $P_{\mathrm{R}}$                  & representative flux: can be $P_{\mathrm{O}}$ or $P_{\mathrm{T}}$ \\
   $\Sigma_{\mathrm{z}}(z)$        &  GRB detector efficiency function with $z$            \\                                 
  $\Sigma_{\mathrm{P}}$        &  GRB detector efficiency function with peak flux            \\                            
  $ \theta_{\mathrm{v}}$  & The viewing angle             \\       
 $\theta_{\mathrm{v,M,J}}$          &  Maximum viewing angle based on the observed flux            \\
  $ \hat{\theta}_{\mathrm{v,M,J}}$  & Most likely $\theta_{\mathrm{v,M,J}}$ given $\Sigma_{\mathrm{P}}$ and $z$              \\      
  $ \hat{\theta}_{\mathrm{v}}$  & Most likely $\theta_{\mathrm{v}}$ based on $\Sigma_{\mathrm{P}}$ and $z$              \\         
  \hline
  \end{tabular}
    \caption{Table of Parameters: Lists the rates, viewing angles, and detector efficiencies used in this study, essential for the analysis presented. Parameters dependent on a structured jet model are indicated by the subscript $\mathrm{J}$.}
  \label{table:rate_parameters}
  \end{center}
\end{table}


\subsection{The relationship between $\Rate_{\mathrm{A,C}}$  and $\Rate_{\mathrm{A,L}}$ }
\label{sub_section_app_and_top_hat}

The relationship between $\Rate_{\mathrm{A,C}}$ from $\Rate_{\mathrm{A,L}}$ for a particular burst can be derived by assuming a dependence on geometry if one assumes a model of the structured jet profile. Given a sample of low-$z$ sGRB detections, this relationship could allow constraints on the jet profile parameter space by future sensitive instruments.

One approach to relate $\Rate_{\mathrm{A,C}}$ with an estimated $\Rate_{\mathrm{A,L}}$ is to extend the framework of \citet{Howell2019MNRAS} which explores the interplay between the maximum viewing angle $\theta_{\mathrm{v,M}}$ and redshift. For a given value of $z$ and a structured jet luminosity profile given as $L(\theta)$, rearranging equation (\ref{eq_max_distance}) we observe a peak flux:

\begin{equation}\label{eq_max_view}
    P(\theta_{\mathrm{v}}) = \frac{L(\theta_{\mathrm{v}}) }{4 \pi D_{\mathrm{L}}(z)^{2}}\,
    \frac{C_{\text{N}}^{-1}}{k(z)}\frac{1}{1 + z}\,.
\end{equation}

\noindent So for a given $z$, a flux value maps to some viewing angle $\theta_{\mathrm{v}}$. Given for some reference value of peak flux $P_{\mathrm{R}}$ there is an associated maximum viewing angle $\theta_{\mathrm{v,M}}$. In the study of \citet{Howell2019MNRAS} a threshold peak flux value was employed as a point estimate of sensitivity, so $P_{\mathrm{R}}=P_{\mathrm{T}}$.  One can could also use the observed value of peak flux, $P_{\mathrm{O}}$, in which $P_{\mathrm{R}}=P_{\mathrm{O}}$, which is more appropriate for calculating $\Sigma_{\mathrm{z}}(z)$ to estimate an apparent rate, $\Rate_{\mathrm{A,C}}$.

\begin{figure}
 \centering
 \includegraphics[scale = 0.65,origin=rl]{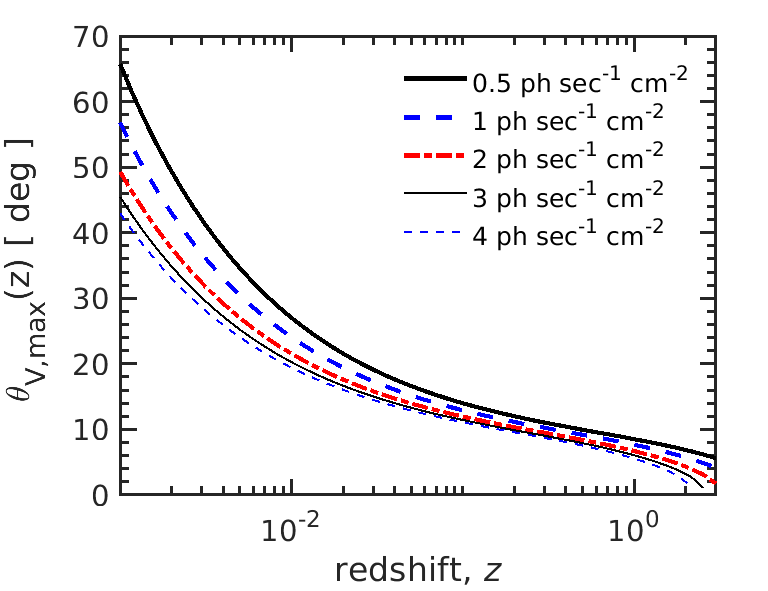}
 \caption{The maximum observable viewing angle as a function of redshift for a sGRB with a structured jet profile based on model A. We assume here 64ms peak fluxes in the 50-300 keV band. The plot shows the interplay between detection threshold and opening angle with increasing redshift. A lower flux sensitivity, that could allow one to probe jet structures at low-$z$, is balanced by decreased detection probability. This is shown for Fermi-GBM in Figure \ref{fig:GBM_Eff_in_50_300keV}.}
  \label{fig:fermi_max_angle_z_flux}
\end{figure}

Figure \ref{fig:fermi_max_angle_z_flux} shows how the maximum viewing angle -- redshift dependence, $\theta_{\mathrm{v,M}}$\,--\,$z$, varies with a range of $P_{\mathrm{R}}$. At low-$z$ there is a greater likelihood of detecting an off-axis event which can artificially boost approximations of the event rate based on volumetric arguments. Thus the need to separate an apparent local rate $\Rate_{\mathrm{A,L}}$ from rates determined at higher-$z$, $\Rate_{\mathrm{A,C}}$.

This effect also impacts the ability to make inferences on the intrinsic rate $\Rate_{\mathrm{I}}$ as the observed properties of the 
jet vary with $z$. This introduces uncertainties into simple scaling assumptions for rates through a standard beaming correction, $(1 - \mathrm{cos}\,\theta_{\mathrm{j}})$, where $\theta_{\mathrm{j}}$ represents an average half-opening angle of the jet. This highlights the need for an understanding of these geometric effects. Such an analysis will enable us to more accurately convert between the observed rates, $\Rate_{\mathrm{A,L}}$ and $\Rate_{\mathrm{A,C}}$, and the intrinsic rate $\Rate_{\mathrm{I}}$, thereby enhancing our understanding of the rate at which events occur. With this in mind we will derive a geometric scaling relation for sGRBs in the following section.
\subsection{A Geometric Scaling Relation for sGRBs}

As the more frequent higher-$z$ population is expected to dominate detection samples, it is crucial to emphasize that these events will likely be observed closer to the jet core. Consequently, it is a reasonable approximation that the viewing angle approximates the core angle $\theta_{\mathrm{c}}$. However, when calculating volumetric rates, the disproportionate influence of nearer events due to geometric bias must be considered for accurate estimates.

To address this, a simple scaling relation is desired to convert an apparent rate  local rate $\Rate_{\mathrm{A,L}}$ from its higher redshift analogue $\Rate_{\mathrm{A,C}}$. Such a scaling will be dependent on $z$ and $P_{\mathrm{R}}$ and can be represented by:

\begin{equation}
\label{eq_geometric_bias_correction}
\eta_{\theta,\mathrm{J}}(z,P_{\mathrm{R}})=
\left\{
\begin{aligned}
     &1 \hspace{1mm}\, \hspace{33mm}\theta_{\mathrm{v,M,J}} \leq \theta_{\mathrm{c,J}}\\
     &\\
     & \frac{1 - \mathrm{cos}\,\theta_{\mathrm{c,J}}}{1 - \mathrm{cos}\, \theta_{\mathrm{v,M,J}}(z,P_{\mathrm{R}})} \hspace{7mm}\, \theta_{\mathrm{c,J}} > \theta_{\mathrm{v,M,J}} \leq \theta_{\mathrm{T}}\\ 
     &\\
          & \frac{1 - \mathrm{cos}\,\theta_{\mathrm{c,J}}}{1 - \mathrm{cos}\,\theta_{\mathrm{T}}} \hspace{21mm}\, \theta_{\mathrm{v,M,J}} > \theta_{\mathrm{T}}
\end{aligned}
\right.
\end{equation}

\noindent where the subscript $\mathrm{J}$ underscores that this relationship relies on a specific structured jet profile; we will adopt this convention for the remainder of this study, in particular for the jet core angle $\theta_{\mathrm{c, J}}$ and the maximum viewing angle $\theta_{\mathrm{v,M,J}}$. This function is derived based on reasonable theoretical assumptions and while reliant on the structured jet profile, introduces a geometric correction across three distinct redshift domains, corresponding to the value of $\theta_{\mathrm{v,M,J}}$. 

At high-$z$, where $\theta_{\mathrm{v,M,J}} < \theta_{\mathrm{c,J}}$, the observable flux predominantly originates from the jet core, rendering the observed rates, $\Rate_{\mathrm{A,L}}$ and $\Rate_{\mathrm{A,C}}$, comparable. For lower $z$ values, contributions from outside the core become significant, with $\theta_{\mathrm{v,M,J}}$ constrained by a truncation angle $\theta_{\mathrm{T}}$, independent of flux sensitivity. We will assume a truncation of order 30\deg as supported by numerical simulations
\citep[i.e.][]{Rezzolla2011ApJ}.

\begin{figure}
 \centering
 \includegraphics[scale = 0.65,origin=rl]{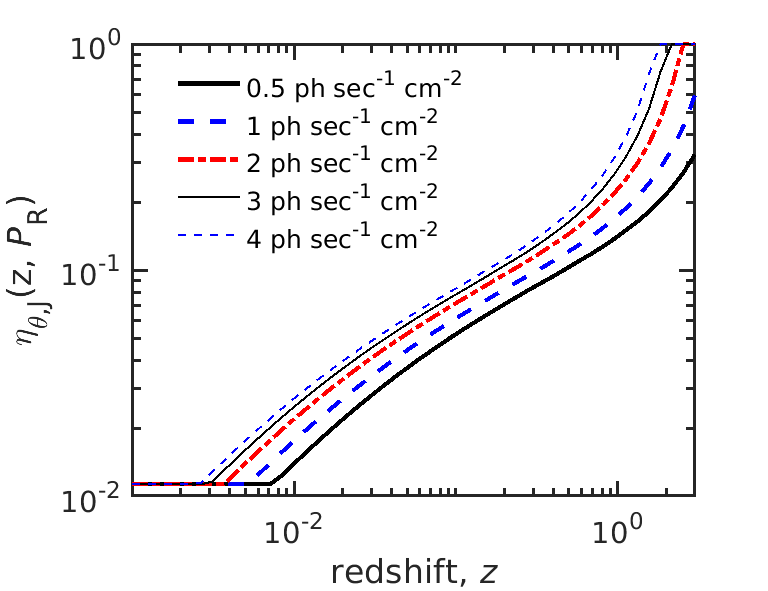}
 \caption{The geometric scaling factor $\eta_{\theta,\mathrm{J}}(z,P_{\mathrm{R}})$ is shown for a range of $z$ for given $P_{\mathrm{L}}$ and using structured jet model A. This factor is a scaling that can be applied to rate estimates obtained from sources with structured jets at low $z$; the proximity means such estimates are artificially boosted by the accessibility to larger viewing angles. At higher-$z$ the effect is minimal and the factor converges towards unity.}
  \label{fig:figure_geometric_bias_correction}
\end{figure}

Figure \ref{fig:figure_geometric_bias_correction} illustrates the correction term $\eta_{\theta,J}$ against redshift for a structured jet profile based on model A, considering various $P_{\mathrm{R}}$ values. At higher-$z$, the correction uniformly equals one, indicating dominance of the jet core in detections. Conversely, at lower-$z$, contributions from the jet's wings are accessible, enhancing the observed rate. We see equation (\ref{eq_geometric_bias_correction}) facilitates scaling of apparent observed rates from off-axis, low-$z$ bursts to a standardized apparent rate $\Rate_{\mathrm{A,C}}$, which is more representative of the sGRB population. This correction diminishes the exaggerated impact of closer events on rate estimates, offering a more balanced understanding when geometric bias is considered.

\begin{figure}
 \centering
 \includegraphics[scale = 0.65,origin=rl]{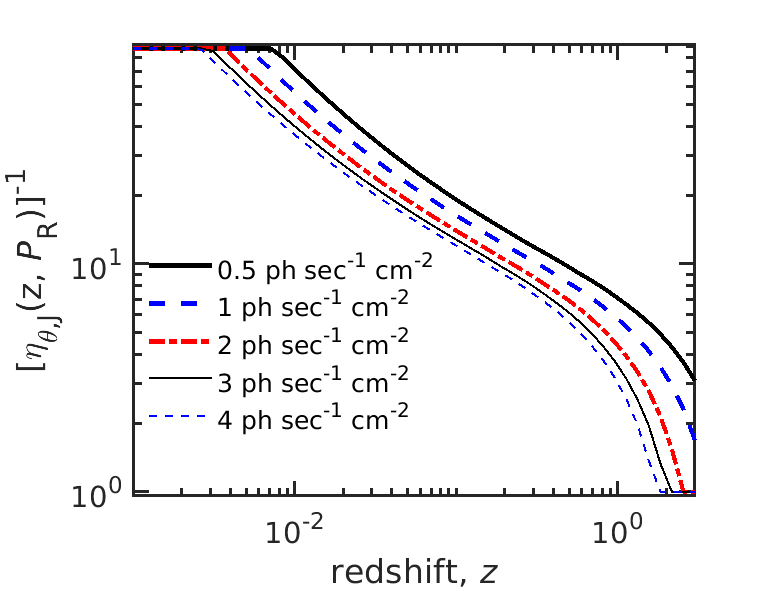}
 \caption{The inverse of $\eta_{\theta,\mathrm{J}}(z,P_{\mathrm{R}})$ shows the actual bias in rate estimation as a function of redshift due to a structured jet profile. We see that a source detected at low-$z$ can have an apparent rate $\Rate_{\mathrm{A,L}}$ orders of magnitude greater than that estimated from the same source observed at higher-$z$. }
  \label{fig:figure_geometric_detection_bias}
\end{figure}

Figure \ref{fig:figure_geometric_detection_bias} is a simple inversion of Figure \ref{fig:figure_geometric_bias_correction} to elucidate the scaling impact of  equation (\ref{eq_geometric_bias_correction}) on an apparent local rate estimate. As an illustration, taking a simple flux point estimate approach at the redshift of GRB~170817A ($z\sim 0.009$), the correction term $1/\eta_{\theta,\mathrm{J}}(z,P_{\mathrm{R}})$, for a standard peak flux value of $P_{\mathrm{R}} \sim 2$ \pflux, is approximately two orders of magnitude based on model A. This significant correction suggests that the inferred rates $\Rate_{\mathrm{A,L}}\sim\mathcal{O}(100)$ \rate, as reported by \citet{ZhangBB2018NatCo, DellaValle2018MNRAS, Salafia2022}, are consistent with estimates of $\Rate_{\mathrm{A,C}}\sim\mathcal{O}(10)$ \rate, as found in \citet{Nakar_2006ApJ,coward_swift_2012,WandermanPiran2015MNRAS}, when the geometric bias is accounted for.

In the following sections, we aim to apply this methodology to GRB~170817A's apparent local observed rate, $\Rate_{\mathrm{A,L}}$, converting it to a cosmological apparent rate $\Rate_{\mathrm{A,C}}$. 

\begin{table*}
  \caption{ The flux, spectral data and redshifts data for the sample of sGRBs observed by Fermi-GBM with secure redshift measurements. All parameters are provided for the 50-300 keV band. Values for the 64ms peak fluxes are shown in photon count and energy units. The spectra models are: power law (pl), exponentially cut-off power law (cpl) and Band
function \citep{Band03}. The spectra are descrobed by a peak energy ($E_{P}$) and low ($\alpha_{S}$) and high ($\beta_{S}$) energy power-law indexs where appropriate. We also show the redshift measurement for each burst.
  Data is taken from the Fermi-GBM burst catalog \citep{vonKienlin_GBM_Burst_Cat4_2020ApJ} at HEASARC: \url{https://heasarc.gsfc.nasa.gov/W3Browse/fermi/fermigbrst.html} }
\begin{center}
  \begin{tabular}{lllllccl}
\hline 
\hline 
GRB & Peak Photon Flux & Peak Energy Flux & Spectrum & $E_{P}$ & $\alpha_{S}$ & $\beta_{S}$ & redshift 
\\ 
& [64ms Ph s$^{-1}$cm$^{2}$] & [64ms erg s$^{-1}$cm$^{2}$ $\times 10^{-7}$] &&&&& 
\\ 
\hline 
170817A & $      2.3 \pm       0.8$ & $      5.2 \pm       1.8$ & cpl & $    229.0\pm      78.0$ & $      0.8\pm       1.4$ & - & $0.0093$
\\ 
211211A & $    201.0 \pm       2.9$ & $    457.0 \pm       6.7$ & band & $   1473.0\pm      76.0$ & $     -0.8\pm       0.0$ & $     -2.5\pm       0.1$ & $0.076$
\\ 
080905A & $      2.4 \pm       0.3$ & $     16.1 \pm       2.0$ & pl & - & $     -1.3\pm       0.1$ & - & $0.12$
\\ 
150101B & $     13.0 \pm       1.4$ & $     28.0 \pm       3.1$ & cpl & $    550.0\pm     190.0$ & $     -0.8\pm       0.2$ & - & $0.13$
\\ 
160821B & $      2.5 \pm       0.3$ & $     12.6 \pm       2.1$ & pl & - & $     -1.6\pm       0.1$ & - & $0.16$
\\ 
150101B & $      1.0 \pm       0.3$ & $      4.8 \pm       1.1$ & pl & - & $     -2.4\pm       0.3$ & - & $0.13$
\\ 
\hline 
  \label{table_sgrb_input_data}
  \end{tabular}
  \end{center}
\end{table*}

\section{The sGRB rate density}
\label{section:sGRB_rate_density}

\subsection{Estimating apparent rates for the Fermi-GBM sGRB sample}
\label{section_sGRB_sample}

Table \ref{table_sgrb_input_data} shows the sample of bursts detected by Fermi-GBM with redshift information. We will use this sample to estimate the apparent local observed rates, $\Rate_{\mathrm{A,C}}$, and show that the rate sample is dominated by GRB~170817A. The table outlines the peak flux and spectral information required to estimate the detection efficiency with redshift $\Sigma_{\mathrm{z}}(z)$ for each burst. We use data from the 50-300 keV detection band for each burst.

Other than for GRB~170817A, the redshift measurements were provided through host galaxy identifications following a precise \emph{Swift} or optical localization. For GRB~170817A we take the distance value of $40.7 +/- 2.4$\,Mpc derived through surface-brightness fluctuation for the distance to NGC 4993 by \citet{Cantiello2018ApJ}. 

Table \ref{table_sgrb_rates_ModelAB} details the calculated rest frame isotropic equivalent luminosities, $L_\mathrm{iso}$, sensitive $\langle VT \rangle$ and volumetric local apparent rates $\Rate_{\mathrm{A,L}}$ for each burst. We note that $L_\mathrm{iso}$ is directly extrapolated using the flux and $z$ value of each burst; therefore for low-$z$ bursts such as GRB~170817A, this represents an apparent luminosity. 

The $\langle VT \rangle$ estimates are determined through equation (\ref{equation_VT2}) and the values of $\Rate_{\mathrm{A,L}}$ are the median 
values and the 90\% credible intervals of the posterior distributions estimated through the framework described in sections \ref{section_rate_Framework} and \ref{section:effect_of_jet_geometry_on_rates}.

The data presented in Tab. \ref{table_sgrb_rates_ModelAB} reveal that the estimates of $\Rate_{\mathrm{A}}$ are overwhelmingly influenced by the closest and, in terms of luminosity, the least bright burst in our sample, GRB~170817A. To illustrate the relative contributions of each burst to the total volumetric rate, we display the maximum detection distances for each burst, calculated with a reference flux limit of 2.2 \pflux which is the value at which detection efficiency is 10\%. For GRB~170817A, this distance reaches a maximum of 42 Mpc. The second nearest burst, GRB211211A, boasts a significantly higher peak flux, making it visible up to the largest distance in our sample, approximately 5.5 Gpc. This burst yields the highest estimate of $\langle VT \rangle$, and consequently, the lowest estimate of $\Rate_{\mathrm{A,L}}$. Given that GRB~170817A predominantly influences the rate calculations, our subsequent analysis will primarily concentrate on this particular burst.

\begin{figure}
 \centering
 \includegraphics[scale = 0.60,origin=rl]{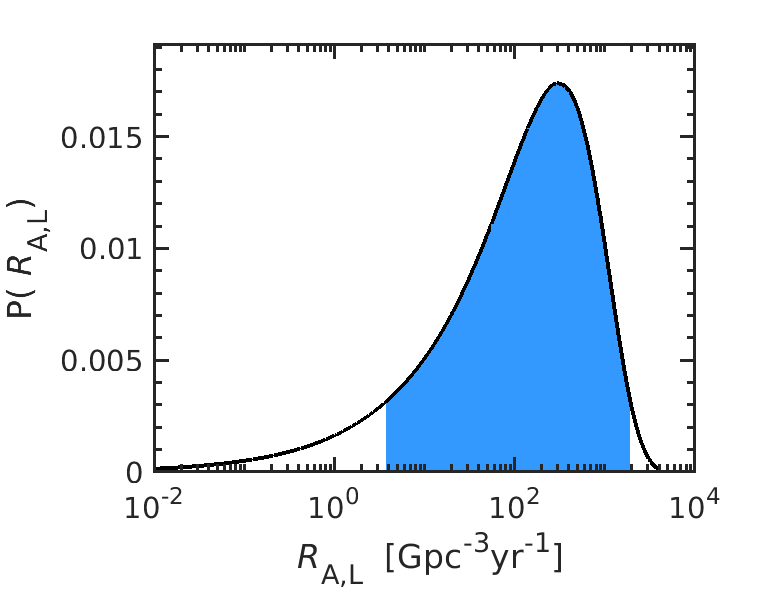}
 \caption{The sGRB apparent local rate posterior based only on the most dominant burst in the Fermi-GBM sample, GRB~170817A. The maximum a posteriori rate and 90\% credible intervals (shown by the shaded region) is $ \Rate_{\mathrm{A,L}}=$\sGRBrateApp \rate.}
  \label{fig:170817A_app_sGRB_posterior}
\end{figure}

Figure \ref{fig:170817A_app_sGRB_posterior} presents the posterior probability distribution for $\Rate_{\mathrm{A,L}}$, centering on the pivotal burst, GRB~170817A. Our analysis yields an apparent sGRB rate of $\Rate_{\mathrm{A}}=$ \sGRBrateApp \rate. A review of Table \ref{table:rate_estimates} demonstrates that our findings are closely aligned with the estimates made by \citet{DellaValle_2018MNRAS,Salafia2022}, despite the utilization of different assumptions as outlined in the table notes. In the next section we will proceed to employ the geometric scaling relation $\eta_{\theta,\mathrm{J}}(z,P_{\mathrm{R}})$ in order to derive a representative value for $\Rate_{\mathrm{A,C}}$.

\subsection{The Apparent Cosmological rate from Fermi-GBM data}

To derive the posterior distribution on $\Rate_{\mathrm{A,C}}$ we apply the same framework as in the previous section, but this time modify equation (\ref{eq_rate_posterior_final}). We apply the inverse geometric scaling relation $1/\eta_{\theta,\mathrm{J}}(z,P_{\mathrm{R}})$ to the function $\langle VT \rangle$ which essentially scales the efficiency function $\Sigma_{\mathrm{z}}(z)$ to account for the redshift dependent effects of beaming:

\begin{equation}
\label{equation_VT_scaled}
\langle VT \rangle= T
 \hspace{-1mm} 
 \int^{\infty}_{0} \frac{\Sigma_{\mathrm{z}}(z) }{\eta_{\theta,\mathrm{J}}(z,P_{\mathrm{R}})} \frac{\mathrm{d}V_{\mathrm{c}} (z) }{\mathrm{d}z } \frac{ 1}{( 1 + z)}\,\mathrm{d}z \,.
\end{equation}
\vspace{1.0mm}

\noindent Here, to ensure a consistent scaling $P_{\mathrm{R}}$ is set to the observed peak flux 2.27 \pflux at each step of $z$.

\begin{figure}
 \centering
 \includegraphics[scale = 0.60,origin=rl]{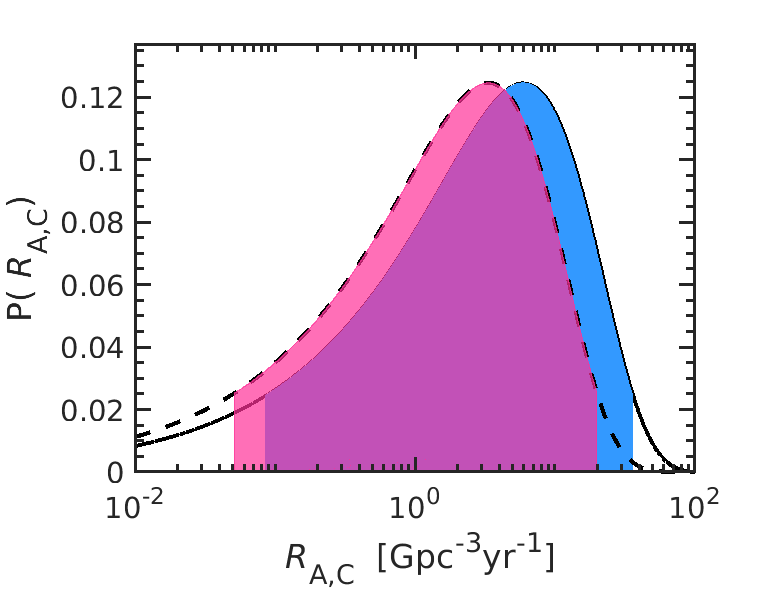}
 \caption{The sGRB apparent cosmological rate posteriors for two structured jet models based only on the most dominant burst in the Fermi-GBM sample, GRB~170817A. The maximum a posteriori rate and 90\% credible intervals are $ \Rate_{\mathrm{A,C}} =$\sGRBrateTopHatA \rate for model A and $ \Rate_{\mathrm{A,C}} =$\sGRBrateTopHatB \rate for model B. }
   \label{fig:170817A_app_cos_sGRB_posterior}
\end{figure}

Figure \ref{fig:170817A_app_cos_sGRB_posterior} presents the posterior probability distributions for $\Rate_{\mathrm{A,C}}$ which yield apparent cosmological rates of $ \Rate_{\mathrm{A,C}} =$\sGRBrateTopHatA\,\rate and $\Rate_{\mathrm{A,C}} =$\sGRBrateTopHatB\,\rate for models A and B respectively. These estimates are interestingly in line with previous estimates based on more distant events observed by $Swift$ \citep{WandermanPiran2015MNRAS, coward_swift_2012}. Given the smaller field-of-view and higher sensitivity of $Swift$, statistically one would expect this instrument to make most detections from the more frequently occurring bursts at relatively higher-$z$ and thus a more face-on orientation. Thus the relative geometrical corrections required for $Swift$ would be less significant.

\begin{table*}
  \caption{The rate related parameters calculated for the sample of Fermi-GBM sGRbs with secure redshifts as shown in table \ref{table_sgrb_input_data}. The first column  shows the sGRB name, $L_\mathrm{iso}$ is the apparent luminosity. The quantity $ \langle V T \rangle$  is the product of the observation time $T$ and the volumetric reach of the search $V$ calculated through equation (\ref{equation_VT2}). A maximum detection distance $D_{\mathrm{L,max}}$, which is highly sensitive to the chosen flux threshold, is provided for illustration using a reference peak flux value 2.2 \pflux which is the value at which the Fermi-GBM detector has a 10\% detection efficiency. The quantity $\Rate_{\mathrm{A,L}}$ is the apparent local rate. The last four columns are model dependent as indicated: the quantity $\Rate_{\mathrm{A,C}}$ is the apparent cosmological rate determined using the framework outlined in section \ref{section:effect_of_jet_geometry_on_rates}; a measure of the maximum viewing angle $\theta_{\mathrm{v,M}}$ is shown for each model for illustration and is calculated using the same reference value of peak flux as used for $D_{\mathrm{L,max}}$.}
\begin{center}
  \begin{tabular}{lllllccll}
\hline 
\hline 
GRB & $L_\mathrm{iso}$ & VT  & $D_{\mathrm{L,max}}$ & $\Rate_{\mathrm{A,L}}$ & $\Rate_{\mathrm{A,C,model A}}$ & $\theta_{\mathrm{v,M,model A}}$  & $\Rate_{\mathrm{A,C,model B}}$ & $\theta_{\mathrm{v,M,model B}}$ 
\\ 
  & [erg s$^{-1}$ ]  & [yr Gpc$^{3}$]  & [Gpc] & [yr$^{-1}$Gpc$^{-3}$] & [yr$^{-1}$Gpc$^{-3}$] & [deg]  & [yr$^{-1}$Gpc$^{-3}$] & [deg]    
\\ 
\hline 
170817A  &  $1.42 \times 10^{47}$  &  $ 1.65 \times 10^{-3}  $  &  $4.16 \times 10^{-2}$  &  $300_{-300}^{+1600}$  &  $5.9_{-5.8}^{+30}$ & $18.78$ &  $2.9_{-2.8}^{+14}$ & $25.74$ 
\\ 
211211A  &  $6.14 \times 10^{51}$  &  $ 2.20 \times 10^{2}  $  &  $3.64$  &  $0.0023_{-0.0013}^{+0.0064}$  &  $0.0023_{-0.0013}^{+0.0064}$ & $8.76$ &  $0.0023_{-0.0013}^{+0.0064}$ & $7.20$ 
\\ 
080905A  &  $1.10 \times 10^{51}$  &  $ 17.70  $  &  $1.15$  &  $0.028_{-0.027}^{+0.093}$  &  $0.006_{-0.005}^{+0.014}$ & $7.06$ &  $0.011_{-0.0096}^{+0.028}$ & $5.17$ 
\\ 
150101B  &  $4.20 \times 10^{50}$  &  $ 40.33  $  &  $1.63$  &  $0.012_{-0.011}^{+0.034}$  &  $0.0036_{-0.0026}^{+0.0091}$ & $8.68$ &  $0.0065_{-0.0055}^{+0.015}$ & $7.11$ 
\\ 
160821B  &  $6.76 \times 10^{50}$  &  $ 28.85  $  &  $1.41$  &  $0.017_{-0.016}^{+0.051}$  &  $0.003_{-0.002}^{+0.008}$ & $7.45$ &  $0.0053_{-0.0043}^{+0.012}$ & $5.55$ 
\\ 
150101B  &  $2.54 \times 10^{50}$  &  $ 6.78  $  &  $0.78$  &  $0.074_{-0.071}^{+0.29}$  &  $0.0077_{-0.0066}^{+0.018}$ & $7.70$ &  $0.0099_{-0.0088}^{+0.025}$ & $5.84$ 
\\ 
\hline 
  \label{table_sgrb_rates_ModelAB}
  \end{tabular}
  \end{center}
\end{table*}

\section{Comparing the rate densities of binary neutron star mergers and sGRBs}
\label{Sect:bns_rates_from_sgrbs}
Prior to GRB~170817A, when GRB rate approximations generally assumed a top-hat geometry, an observed rate $\Rate_{\mathrm{A,C}}$ was often related to an intrinsic rate $\Rate_{\mathrm{I}}$ by the scaled relation $(1 - \mathrm{cos}\,\theta_{\mathrm{j}})$ where $\theta_{\mathrm{j}}$ is the average observed jet half-opening angle. For a jet to produce an sGRB, it must first successfully form and then propagate through the merger ejecta. This depends on a complex interplay of factors including the post-merger remnant \citep{Kiuchi2023} density and structure of the ejecta \citep{Bromberg_2012ApJ, Lazzati2017, Hamidani2020}, and the jet's intrinsic properties \citep{Gottlieb2023ApJ, Kathirgamaraju2019}. Therefore, this a relation can also provide insights on the fraction of successful jets assuming that $\Rate_{\mathrm{I}}=\Rate_{\mathrm{BNS}}$ \citep{coward_swift_2012,Sarin2022PhRvD,Escorial2023ApJ}.

Given that most sGRBs are observed at distances in which wider angled emissions are not accessible \citep{OConnor2024}, a top-hat approach can allow reasonable approximations. However, in the strictest sense, for detailed modeling, a top-hat approximation does not account for the observational dependency of a given jet profile with $z$ as emphasized by Fig.\ref{fig:170817_eff_with_z}. This latter fact will be key when future sensitive instruments are able to accumulate more low-$z$ events viewed at wider angles.

Without access to a standard value of $\theta_{\mathrm{j}}$ the methodology presented in this paper can provide a different approach to estimate of the successful jet fraction. Under a SJ scenario, the maximum viewing angle $\theta_{\mathrm{v,M,J}}$ can be a surrogate for $\theta_{\mathrm{j}}$ as it naturally incorporates a $z$ dependency. 

As an illustration, if we assume here that the fraction of successful jets is 100\%, then we can derive a representative BNS rate, $\Rate_{\mathrm{BNS}}'$, through:

\begin{equation}
\Rate_{\mathrm{BNS}}' =  \Rate_{\mathrm{A,C}} \cdot [1 - \cos(\theta_{\mathrm{v,M,J}}(z, P_{\mathrm{R}})]^{-1}\,.
\end{equation}

\noindent Here the variable $\Rate_{\mathrm{A,C}}$ is transformed by scaling it with a constant factor $ [1 - \cos(\theta_{\mathrm{v,M,J}}(z, P_{\mathrm{R}})]^{-1}$. Therefore the posterior distribution, $ P(\Rate_{\mathrm{BNS}}') $ must be obtained from
 $ P(\Rate_{\mathrm{A,C}}) $ through a transformation of variables. If we set $\Lambda_{SF} =  [1 - \cos(\theta_{\mathrm{v,M,J}}(z, P_{\mathrm{R}})]^{-1}$ then: 

\begin{equation}
P(\Rate_{\mathrm{BNS}}') = P\left(\frac{\Rate_{\mathrm{A,C}}}{\Lambda_{SF}}\right) \times \frac{1}{|\Lambda_{SF}|}
\end{equation}

\noindent where the Jacobian of the transformation, $1/ |\Lambda_{SF}|$ modifies $ P(\Rate_{\mathrm{A,C}}) $ to account for the change in scale ensuring that the total probability remains equal to unity. The absolute value is used to ensure the scale factor is positive. 

As a demonstration we take the maximum viewing angle of both model-A and model-B at a representative redshift given by the median value of the GBM redshift sample provided in Table \ref{table_sgrb_input_data}, $z_{\mathrm{R}} = 0.1$. Using a reference peak flux, $P_{\mathrm{R}}=$ 2.2 \pflux, the 10\% efficiency value of GBM as used in Table \ref{table:rate_estimates}, we find, using in Eq.~\ref{eq_max_view}, values of $\theta_{\mathrm{v,M,A}}$ = 10.8\deg and $\theta_{\mathrm{v,M,B}}$ = 9.3\deg. 

\begin{figure}
 \centering
 \includegraphics[scale = 0.60,origin=rl]{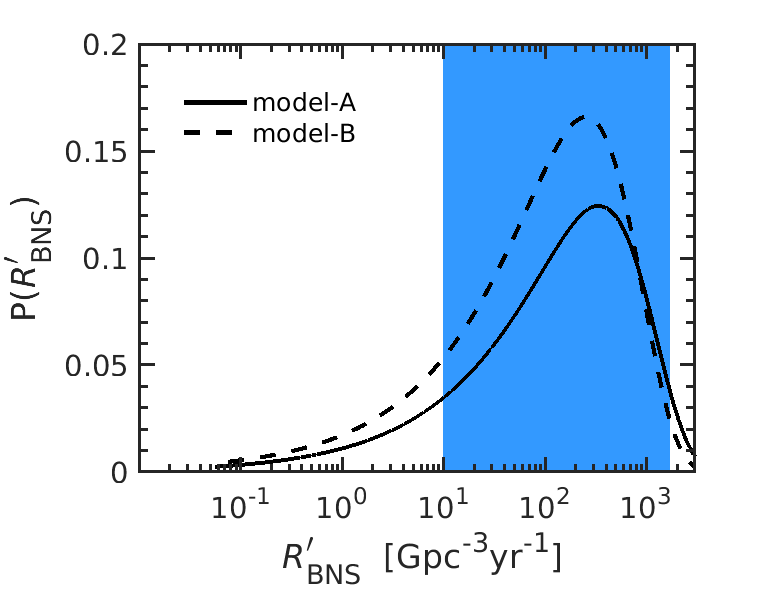}
 \caption{The representative BNS rate posteriors $P(\Rate_{\mathrm{BNS}}')$ based on the two structured jet models A and B. These are obtained by applying a transformation to the sGRB apparent cosmological rate posteriors $ P(\Rate_{\mathrm{A,C}})$ of Fig \ref{fig:170817A_app_cos_sGRB_posterior}. The BNS merger rates constraints of \citet{2021_GWTC3_cat} are indicated by the shaded area.}
   \label{fig:BNS_from_sGRB_Posterior}
\end{figure}

Figure \ref{fig:BNS_from_sGRB_Posterior} shows the representative posterior distributions of $\Rate_{\mathrm{BNS}}'$ using $\theta_{\mathrm{v,M,A}}$ and $\theta_{\mathrm{v,M,B}}$. We find rates of $\Rate_{\mathrm{BNS}}'=340_{-330}^{+1700}$ \rate and $\Rate_{\mathrm{BNS}}'=250_{-250}^{+1300}$ \rate for model-A and model-B respectively. For comparison with estimated BNS rates the shaded region shows the $\Rate_{\mathrm{BNS}}$ range of  $10-1700$ \rate from \citet{2021_GWTC3_cat}. We suggest that this methodology can be used to draw comparisons between sGRB and BNS merger rates but also to obtain estimates of the fraction of successful sGRB jets.

\section{Conclusions}
\label{section:conclusions}
In this study we have reconciled the discrepancies between the apparent rate estimates of short gamma-ray bursts from the singular event GRB~170817A and predictions based on pre-2017 top-hat jet models. By incorporating Fermi-GBM detection efficiency and structured jet profiles into our analysis, we have developed a comprehensive framework that accurately accounts for observational biases introduced by jet geometry at low redshifts.

Our results emphasize the importance of geometric corrections in understanding the observed discrepancies in sGRB event rates. Specifically, we have shown that elevated sGRB event rates from sources at low-$z$ are primarily driven by geometrical effects. Once the geometric biases are understood and removed, the rates align with previous population rates for sGRBs obtained from high-$z$ sources. We suggest that under a structured jet scenario, event rates determined from low-$z$, $\Rate_{\mathrm{A,L}}$, are apparent due to geometry and rates determined from higher-$z$, $\Rate_{\mathrm{A,C}}$ are apparent due to beaming effects. 

As a demonstration, we have determined an apparent low-$z$ rate of GRB~170817A-like events of $ \Rate_{\mathrm{A,L}} =$\sGRBrateApp\,\rate. When we recalculate the rates, adjusting for geometrical effects, we find apparent cosmological sGRB rates of $ \Rate_{\mathrm{A,C}} =$\sGRBrateTopHatA\,\rate and $\Rate_{\mathrm{A,C}} =$\sGRBrateTopHatB\,\rate for models A and B respectively. These numbers are in line with other studies of high-$z$ sGRBs \citep{Nakar_2006ApJ, coward_swift_2012, WandermanPiran2015MNRAS, Mandel2022LRR}. 

Notably, our proposed framework is model-dependent, emphasizing that a more confident understanding of rates can help discriminate between different structured jet models. This model dependency is crucial as it suggests that enhancements in our understanding of detection efficiencies and jet geometries could lead to more refined models that accurately predict the occurrence rates of sGRBs. It is a logical next step to test the scaling relations in population inference of the rate and intrinsic luminosity function of sGRBs.

An important aspect of our study is the comparison between estimating sGRB rates using detection efficiencies and point estimates. We have demonstrated that using detection efficiency functions, rather than single flux threshold point estimates, provides a more accurate and nuanced approach. We have shown how rate estimates are highly sensitive to the chosen flux threshold. Modelling a detector efficiency function allows for the incorporation of real observational constraints and variability in detector sensitivities, thereby offering a substantial improvement over simpler models.

Our study suggests that one cannot exactly convert a rate $\Rate_{\mathrm{A,C}}$ to an intrinsic rate $\Rate_{\mathrm{I}}$ as an average beaming factor does not truly exist; the geometric scaling is both redshift and structural jet model dependent. Without access to a standard value of $\theta_{\mathrm{j}}$ we suggest that the maximum viewing angle $\theta_{\mathrm{v,M,J}}$ could be a surrogate for $\theta_{\mathrm{j}}$ as it naturally incorporates a $z$ dependency. We find that this approach is able to demonstrate some parity between $\Rate_{\mathrm{A,C}}$ and a BNS rate $\Rate_{\mathrm{BNS}}$.


In conclusion, this study underlines the necessity of considering jet geometry and detection biases when estimating rates of astronomical phenomena like sGRBs. Our methodologies enhance our understanding of these events and set the stage for future observational strategies and theoretical developments in $\gamma$-ray burst astrophysics. This refined approach not only facilitates accurate interpretation of apparent rate estimates but also provides insights into the potential to discriminate between jet models, contributing to a deeper understanding of the mechanisms driving these extraordinary cosmic events.


\appendix

\section{The effect of the angular timescale on the luminosity of a structured jet}
\label{app:angular_time_scale}
\begin{figure}
 \centering
 \includegraphics[scale = 0.65,origin=rl]{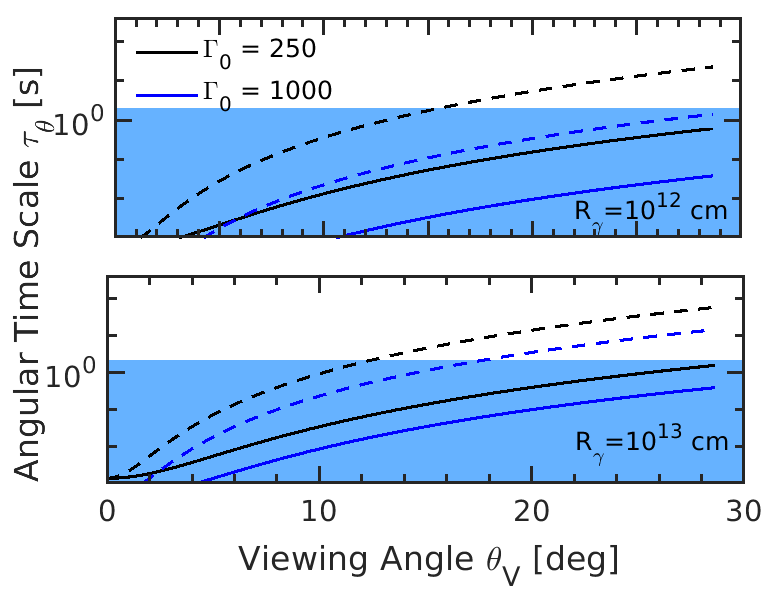}
 \caption{A toy model shows the effect of the angular time scale $\tau_{\theta} $ with viewing angle $\theta_{\text{V}}$. The shaded area represents the observed GRB duration set as $ \tau_{\text{GRB}} =2$s. The solid curves represent a structured jet core parameter of $\theta_c=5$ and the dashed curves, $\theta_c=2$. The colors of the curves represent different values of $\Gamma_0$ and the upper and lower panels are based on the indicated values of the $\gamma$-ray emitting radius, $R_\gamma$. The plots show that $\tau_{\theta} $ becomes larger with with $\theta_{\text{V}}$ leading to increased pulse overlap. This effect is enhanced with smaller $\theta_c$ and larger $R_\gamma$.
  \label{fig:ang_time_VS_theta_v}}
\end{figure}

To consider the effect of pulse overlap on GRB luminosity, one must consider the angular timescale \( t_{\text{ang}} \), which represents the time difference between the arrival of photons emitted from different parts of the jet due to the jet's geometry and relativistic effects. The variability from internal shocks, magnetic reconnections, or other dynamic processes within the jet leads to a series of short emission pulses. Pulse overlap occurs when multiple emission pulses overlap in time as observed from a given viewing angle. The degree of overlap is influenced by both the intrinsic variability of the jet and the angular timescale.

Figure \ref{fig:ang_time_VS_theta_v} shows a toy model to examine the effect on variations in $\Gamma_0$ and $R_\gamma$. We compute \( T_{\theta} \) for two values of $R_\gamma$ [$10^{12}$\,cm, $10^{13}$\,cm], the latter corresponding to GRB~170817A and model the Lorentz factor as a function of \( \theta_{\text{V}} \) through $\Gamma(\theta) = \Gamma_0 / (1 + (\theta / \theta_c)^2)$ so that:

\[
\tau_{\theta}(\theta_{\text{V}}) = \frac{R_\gamma }{ c \Gamma_0^{2}} [ 1 + (\theta_{\text{V}} / \theta_c)^2  ]^{2}\,.
\]
\noindent We further set two values of both $\Gamma_0$ [250, 1000] and 
$\theta_c$ [2\deg, 5\deg].

The plots show that to the jet core, \( \tau_{\theta} \) will be small relative to the intrinsic duration of the pulses; thus the pulses will appear more distinct and overlap less allowing pronounced variability in the observed light curve. When \( \tau_{\theta} \) is large at wider angles, pulses emitted from different parts of the jet will overlap more in time. This overlap can smooth out the observed light curve, reducing the apparent variability. One can see that the latter effect is more pronounced with lower $\Gamma_0$, higher values of $R_\gamma$ and a more compact value of $\theta_c$.


\bibliographystyle{mnras}
\bibliography{sgrb_rates}

\end{document}